\documentclass[aps,showpacs,nofootinbib,superscriptaddress,twocolumn]{revtex4}

\usepackage{graphicx}
\usepackage{bm}
\usepackage{amsmath}
\usepackage{amssymb}
\usepackage{hyperref}

\newcommand{\xbj}{x}

\newcommand{\qslash}{\kern 0.2 em n\kern -0.50em /}
\newcommand{\nslash}{\kern 0.2 em n\kern -0.50em /}
\newcommand{\kslash}{\kern 0.2 em k\kern -0.45em /}
\newcommand{\lslash}{\kern 0.2 em l\kern -0.50em /}
\newcommand{\pslash}{\kern 0.2 em p\kern -0.50em /}
\newcommand{\Sslash}{\kern 0.2 em S\kern -0.50em /}
\newcommand{\Pslash}{\kern 0.2 em P\kern -0.50em /}
\newcommand{\Dslash}{\kern 0.2 em D\kern -0.65em /\kern 0.15em}

\newcommand{\bp}{\boldsymbol{p}_T}
\newcommand{\bP}{\boldsymbol{P}_T}
\newcommand{\bk}{\boldsymbol{k}_T}

\newcommand{\ssh}{\!\!\!/}
\newcommand{\Tr}{\operatorname*{Tr}\nolimits}

\newcommand{\ph}{\phi_h}

\begin{document}

\title{ Double spin asymmetries $A_{LT}^{\cos\phi_S}$ and $A_{LT}^{\cos(2\phi_h -\phi_S)}$ in semi-inclusive DIS}

\author{Wenjuan Mao}\affiliation{Department of Physics, Southeast University, Nanjing
211189, China}
\affiliation{School of Physics and State Key Laboratory of Nuclear Physics and Technology, Peking University, Beijing
100871, China}
\author{Zhun Lu}\email{zhunlu@seu.edu.cn}\affiliation{Department of Physics, Southeast University, Nanjing
211189, China}
\author{Bo-Qiang Ma}\email{mabq@pku.edu.cn}\affiliation{School of Physics and State Key Laboratory of Nuclear Physics and Technology, Peking University, Beijing
100871, China}
\affiliation{Collaborative Innovation Center of Quantum Matter, Beijing 100871, China}
\affiliation{Center for High Energy Physics, Peking University, Beijing 100871, China}

\author{Ivan Schmidt}\email{ivan.schmidt@usm.cl}
\affiliation{Departamento de F\'\i sica, Universidad T\'ecnica Federico Santa Mar\'\i a, and
Centro Cient\'ifico-Tecnol\'ogico de Valpara\'iso,
Casilla 110-V, Valpara\'\i so, Chile}

\begin{abstract}
We investigate the double spin asymmetries of pion production in semi-inclusive deep inelastic scattering with a longitudinal polarized beam off a transversely polarized proton target.
Particularly, we consider the $\cos\phi_S$ and $\cos(2\phi_h -\phi_S)$ modulations, which can be interpreted by the convolution of the twist-3 transverse momentum dependent distributions and twist-2 fragmentation functions.
Three different origins are taken into account simultaneously for each asymmetry: the $g_T D_1$ term, the $e_T H_1^\perp$ term, and the $e_T^\perp H_1^\perp$ term in the $\cos\phi_S$ asymmetry; and the $g_T^\perp D_1$ term, the $e_T H_1^\perp$ term, and the $e_T^\perp H_1^\perp$ term in the $\cos(2\phi_h -\phi_S)$ asymmetry.
We calculate the four twist-3 distributions $g_T(x,\bm k_T^2)$, $g_T^\perp(x,\bm k_T^2)$, $e_T(x,\bm k_T^2)$, and $e_T^\perp(x,\bm k_T^2)$ in a spectator-diquark model including vector diquarks.
Then we predict the two corresponding asymmetries for charged and neutral pions at the kinematics of HERMES, JLab, and COMPASS for the first time. The numerical estimates indicate that the two different angular-dependence asymmetries are sizable by several percent at HERMES and JLab, and the $\cos\phi_S$ asymmetry has a strong dependence on the Bjorken $x$.
Our predictions also show that the dominant contribution to the $\cos\phi_S$ asymmetry comes from the $g_T D_1$ term, while the $g_T^\perp D_1$ term gives the main contribution to the $\cos(2\phi_h -\phi_S)$ asymmetry; the other two $T$-odd terms almost give negligible contributions. In particular, the $\cos(2\phi_h -\phi_S)$ asymmetry provides a unique opportunity to probe the distribution $g_T^\perp$.

\end{abstract}

\pacs{12.39.-x, 13.60.-r, 13.88.+e}

\maketitle

\section{Introduction}

In the last two decades, azimuthal asymmetries in spin-polarized semi-inclusive deeply inelastic scattering (SIDIS) in the small transverse momentum region have been explored extensively by experimental and theoretical studies (for reviews, see~\cite{bdr,D'Alesio:2007jt,Barone:2010ef,Boer:2011fh}).
Of particular interest are the Sivers asymmetry~\cite{Sivers:1989cc,Anselmino:1994tv,Brodsky:2002cx} and the Collins asymmetry~\cite{Collins:1992kk}, which have been measured by the HERMES Collaboration~\cite{Airapetian:2009ae,
Airapetian:2010ds}, the COMPASS Collaboration~\cite{Alexakhin:2005iw,Ageev:2006da,Alekseev:2008aa,
Alekseev:2010rw,Adolph:2012sn,Adolph:2012sp}
and the Jefferson Lab (JLab) Hall A Collaboration~\cite{Qian:2011py,Zhao:2014qvx}. These asymmetries provide great opportunities to access novel distributions of unpolarized quark/hadron inside a transversely polarized nucleon/quark, and therefore they are crucial for the understanding of the transverse spin and momentum structure of nucleon.
Recently, further asymmetries beyond the Sivers and Collins asymmetries also receive growing attention, such as the $\sin(3\phi_h-\phi_S)$ asymmetry~\cite{Zhang:2013dow} that involves the pretzelosity distribution $h_{1T}(x,\bm k_T^2)$~\cite{Avakian:2008dz,She:2009jq,Lorce:2011kn}, and the $\cos(\phi_h-\phi_S)$ double spin asymmetry~\cite{Huang:2011bc} contributed by $g_{1T}(x,\bm k_T^2)$~
\cite{Kotzinian:2006dw,Boffi:2009sh,Zhu:2011zza}. These are leading-twist asymmetries.
On the other hand, measurements of several single spin asymmetries (SSAs) appearing at subleading-twist level, i.e., the longitudinally beam spin asymmetry $A_{LU}^{\sin\phi_h}$~\cite{clas04,hermes07,Aghasyan:2011ha,Gohn:2014zbz} and the longitudinal target spin asymmetry $A_{UL}^{\sin\phi_h}$~\cite{Airapetian:2005jc,Avakian:2013sta}, were also performed.
Sizable asymmetries have been observed and have provided the basis for several related theoretical studies~\cite{Bacchetta:2004jz,Metz:2004je,Song:2010pf,
Mao:2012dk,Song:2013sja,Song:2014sja,Lu:2014fva}. Here we should mention a recent theoretical prediction~\cite{Mao:2014aoa} on the transverse SSAs at subleading twist.
These asymmetries are of vital importance, as they provide complementary information on the spin and flavor structure of nucleon.

Encouraged by the sizable asymmetries in SSAs at twist-3 level, in this work we will consider the case of double polarized SIDIS, in which a longitudinally polarized lepton beam collides on a transversely polarized nucleon target.
Except for the aforementioned $\cos(\phi_h-\phi_S)$ asymmetry that appears at leading twist, theoretically there are two other angular modulations (assuming one photon exchange), the $\cos\phi_S$ and the $\cos(2\phi_h-\phi_S)$ moments, which may also receive nonvanishing contributions.
As shown in Ref.~\cite{Bacchetta:2006tn}, by assuming tree-level TMD factorization, each of the two double spin asymmetries (DSAs) can be interpreted as the convolution of twist-3 transverse momentum dependent (TMD) distributions and fragmentation functions (FFs) and their twist-2 counterparts.
Since there are less systematic studies and calculations on the $\cos{\phi_S}$ and $\cos{(2\phi_h-\phi_S)}$ asymmetries in the literature to reveal the related transverse spin structure of the nucleon at twist 3, our main purpose is to give a phenomenological study on the feasibility of experimental measurements on these transverse target DSAs at subleading twist.
Particularly, we will focus on the roles of twist-3 TMD distributions in DSAs by applying the Wandzura-Wilczek approximation~\cite{Wandzura:1977qf} to ignore the contribution from interaction-dependent twist-3 FFs.

In this scenario, there are four twist-3 TMD distributions involved in the transverse target DSAs: $g_T(x,\bm k_T^2)$, $g_T^\perp(x,\bm k_T^2)$, $e_T(x,\bm k_T^2)$, and $e_T^\perp(x,\bm k_T^2)$.
Among them, $g_T$ contributes to the $\cos\phi_S$ asymmetry, while $g_T^\perp$ contributes to the $\cos{(2\phi_h-\phi_S)}$ asymmetry; $e_T$ and $e_T^\perp$ contribute to both asymmetries through the convolution with the Collins FF $H_1^\perp$.
Among these distributions, $g_T(x,\bm k_T^2)$, especially its integrated version $g_T(x)$~\cite{Jaffe:1990qh}, has been studied extensively in literature~\cite{Mulders:1995dh,Jakob:1997wg,Metz:2008ib,Accardi:2009au,Efremov:2009ze}.
The other $T$-even distribution $g_T^\perp$ has been calculated in the spectator-diquark model~\cite{Jakob:1997wg} and the bag model~\cite{Avakian:2010br}.
The $T$-odd and chiral-odd distribution $e_T$ was proposed in Ref.~\cite{Boer:1997nt}, while another $T$-odd distribution $e_T^\perp$ was introduced in Ref.~\cite{Goeke:2005hb}.
The two $T$-odd distributions can be viewed as the analogy of the Sivers function at twist-3 level, and have been studied in Refs.~\cite{Gamberg:2006ru,Lu:2012gu} in scalar-diquark models.
We note that sizable transverse spin asymmetries of charged pion production related to twist-3 dynamics are being measured at COMPASS~\cite{Parsamyan:2013ug,Parsamyan:2014uda}, and it may be quite interesting and necessary to give theoretical estimates for further comparisons.

The remaining content of this paper is organized as follows.
In Sec.~II, we will calculate the four TMD distributions $g_T$, $g_T^\perp$, $e_T$, and $e_T^\perp$ for the $u$ and $d$ valence quarks, as it is necessary to know their magnitudes and signs to predict the transverse target DSAs.
We will apply a spectator-diquark model, which was also used in Refs.~\cite{Bacchetta:2008af,Mao:2012dk,Mao:2013waa,Mao:2014aoa,Liu:2014vwa}.
In Sec.~III, we will present our predictions on the $\cos\phi_S$ and $\cos(2\phi_h -\phi_S)$ asymmetries for charged and neutral pions in SIDIS, using the model results obtained in Sec.~II.
In this calculation, we consider the experimental configurations accessible at HERMES, JLab, and COMPASS.
Although the TMD factorization at twist-3 level has not been proved~\cite{Gamberg:2006ru,Bacchetta:2008xw}, here we would like to adopt a more phenomenological approach, i.e., to use the tree-level result in Ref.~\cite{Bacchetta:2006tn} to perform the estimate.
Finally, we summarize this work in Sec.~IV.

\section{Model calculation of twist-3 TMD distributions}
\label{functions}
\begin{figure}
  \includegraphics[width=0.8\columnwidth]{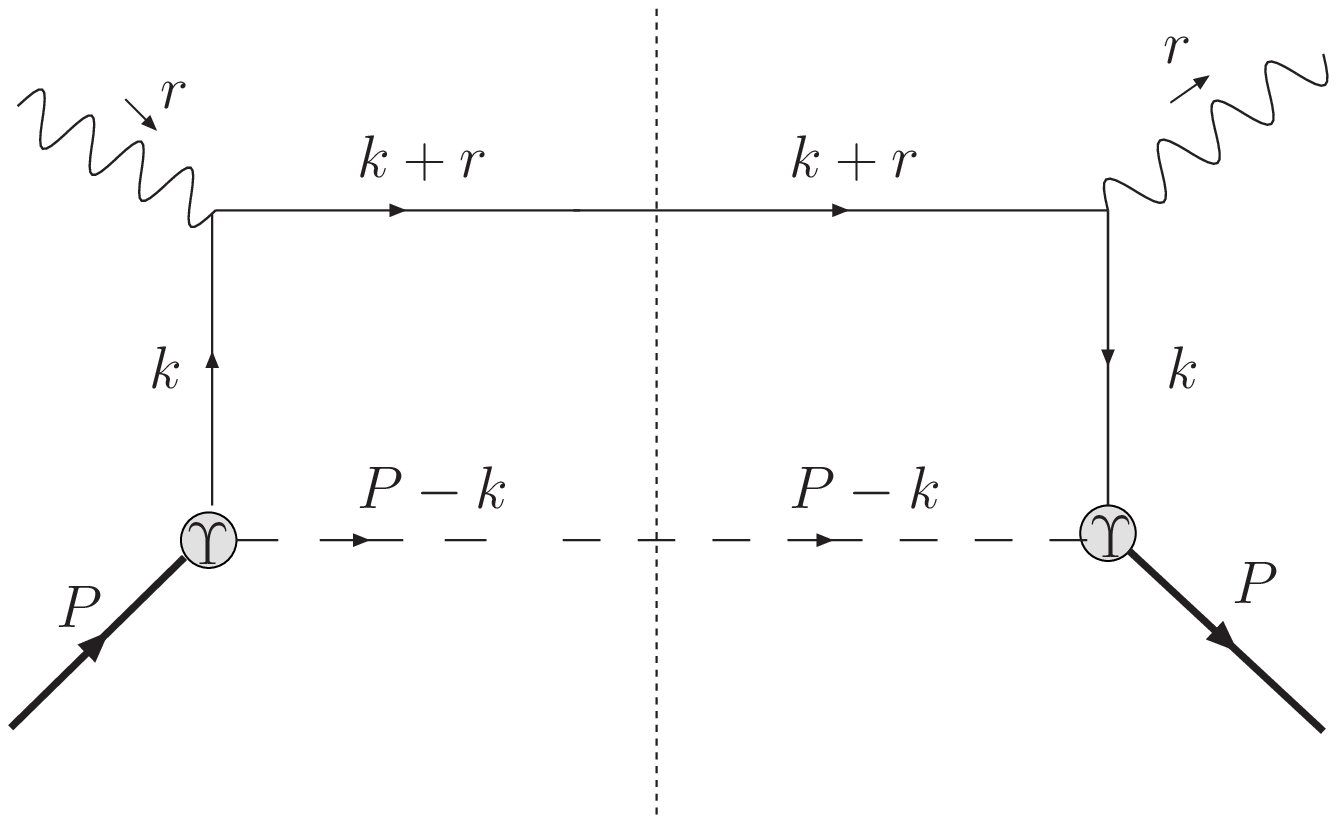}\\
  \includegraphics[width=0.8\columnwidth]{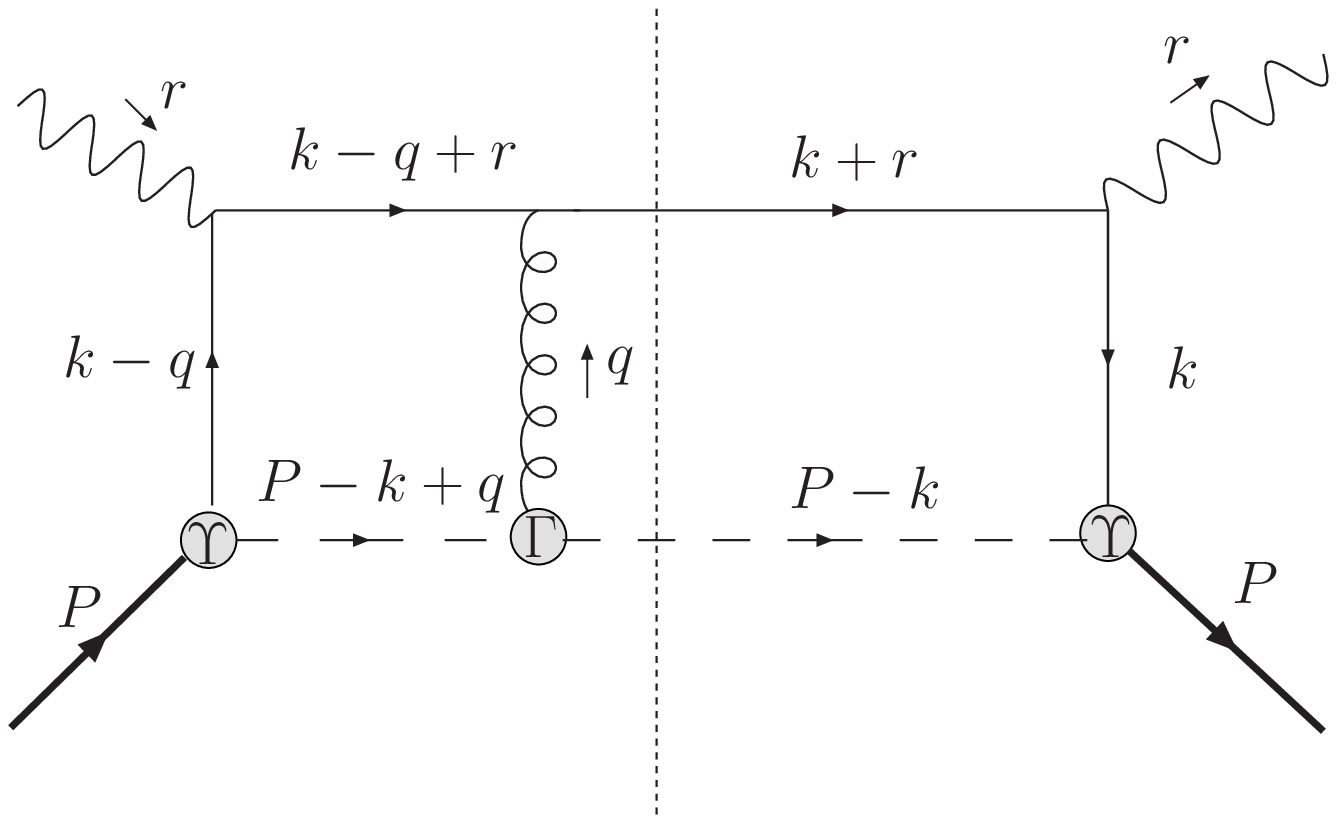}
 \caption{Cut diagrams for the spectator model calculation at tree level (upper) and one-loop level (lower).
 The dashed lines denote the spectator diquarks, which can be scalar diquarks or axial-vector diquarks.}
 \label{diagrams}
\end{figure}

In this section, we calculate the twist-3 TMD distributions $g_T(x,\bm k_T^2)$, $g_T^\perp(x,\bm k_T^2)$, $e_T(x,\bm k_T^2)$, and $e_T^\perp(x,\bm k_T^2)$ within the framework of spectator-diquark model.
To obtain the TMD distribution for the $u$ and $d$ quarks, we need to consider the contributions from both the scalar diquarks and vector diquarks.
For the former one, we will apply the scalar-diquark model which has been widely used to calculate the TMD distributions~\cite{Brodsky:2000ii,Brodsky:2002cx,jy02,Boer:2002ju,Gamberg:2006ru,Lu:2012gu}.
To include the contribution from the vector diquarks, we will use the approach developed in Ref.~\cite{Bacchetta:2008af}, that is, to adopt the light-cone polarization sum for the vector diquark and use a general relation between quark flavors and diquark types.

We start from the expression of the gauge-invariant quark-quark correlator,
\begin{align}
\Phi(x,\bm k_T)&=\int {d\xi^- d^2\bm{\xi}_T\over (2\pi)^3}e^{ik\cdot\xi}
\langle PS|\bar{\psi}_j(0)\mathcal{L}[0^-,\infty^-]\nonumber\\
& \times \mathcal{L}[\bm 0_T,\bm \xi_T]\mathcal{L}[\infty^-,\xi^-]\psi_i(\xi)|PS\rangle\,.
\label{Phi}
\end{align}
Here $\mathcal{L}$ is the future pointing gauge link, corresponding to the SIDIS process, and $k$ and $P$ are the momenta of the struck quark and the target nucleon, respectively.
At the twist-3 level, the correlator (\ref{Phi}) for a transversely polarized nucleon can be decomposed into
\begin{align}
\Phi(x,\bm k_T,\bm S_T)\bigg{|}_{\textrm{twist-3}}
&= {M\over 2P^+}\left\{\gamma^\alpha\gamma_5 \left(S_T^\alpha g_T^\prime +\frac{\bm k_T\cdot \bm S_T}{M^2} k_T^\alpha g_T^\perp\right)\right.\nonumber\\
+&\left.i\gamma_5\frac{\bm k_T\cdot \bm S_T}{M} e_T-\frac{\epsilon_T^{\rho\sigma}k_{T\rho}S_{T\sigma}}{M} e_T^{\perp}+\cdots\right\},\label{phist}
\end{align}
where $\cdots$ denotes the other twist-3 distributions that are not considered in this work.
For convenience, we adopt the light-front coordinates $a^{\pm}=(a^0\pm a^3)/\sqrt{2}=a\cdot n_{\mp}$ for an arbitrary four-vector $a=[a^+,a^-,\bm a_T]$, where the two lightlike vectors are defined as $n_+=[0,1,\bm{0}_T]$ and $n_-=[1,0,\bm{0}_T]$.

Obviously, the four related twist-3 TMD distributions $g_T$, $g_T^\perp$, $e_T$, and $e_T^\perp$ can be obtained from the correlator $\Phi(x,\bm k_T,\bm S_T)$ by the following traces:
\begin{align}
\frac{1}{2}\Tr[\Phi\gamma^\alpha\gamma_5]&=\frac{M}{P^+}\left[S_T^\alpha g^\prime_T+ \frac{\bm k_T\cdot\bm S_T}{M^2}k_T^\alpha g_T^{\perp}\right],
\label{gTs}\\
\frac{1}{2}\Tr[\Phi i \gamma_5]&=\frac{M}{P^+}\left[\frac{\bm k_T\cdot \bm S_T}{M}e_T\right], \label{eT}\\
\frac{1}{2}\Tr[\Phi]&=\frac{M}{P^+}\left[-\frac{\epsilon_T^{\rho\sigma} k_{T\rho}S_{T\sigma}}{M} e_T^\perp\right].\label{eTperp}
\end{align}
In the trace of Eq.~(\ref{gTs}) we have applied the notation in Refs.~\cite{Mulders:1995dh,Goeke:2005hb}.
One can obtain the expression for $g_T$ from $g_T^\prime$ and $g_T^\perp$ via the combination~\cite{Bacchetta:2006tn}
\begin{align}
g_T(x,\bk^2)=g_T^\prime(x,\bk^2)+\frac{\bm k_T^2}{2M^2}g_T^\perp(x,\bk^2).
\end{align}

In the spectator models~\cite{Jakob:1997wg,Bacchetta:2003rz,Gamberg:2007wm,Bacchetta:2008af}, the relevant diagrams used to calculate the correlator (\ref{Phi}) from the scalar diquark and the axial-vector diquark  are shown in Fig.~\ref{diagrams}.
In the lowest-order expansion of the gauge link, which means setting $\mathcal{L}=1$, we apply the upper panel of Fig.~\ref{diagrams} to obtain the correlators ${\Phi_s}^{(0)}$ and ${\Phi_v}^{(0)}$ that are contributed by the scalar and the axial-vector diquarks, respectively, as
\begin{align}
\Phi^{(0)}_s(x,\bk)&\equiv \frac{N_s^2(1-x)^3}{32 \pi^3 P^+}\frac{\left[ (k\ssh +m)\gamma_5 S\ssh (P\ssh +M) (k\ssh +m)\right]}{(\bk^2+L_s^2)^4}, \label{lophis}\\
 \Phi^{(0)}_{v}(x,\bk)&\equiv \frac{N_v^2(1-x)^3}{64 \pi^3 P^+}d_{\mu\nu}(P-k)\nonumber\\
&\times \frac{\left[(k\ssh +m)\gamma^{\mu}\gamma_5 S\ssh(M-P\ssh )\gamma^{\nu}(k\ssh+m)\right]}{(\bk^2+L_v^2)^4},
\label{lophiv}
\end{align}
where $N_s$ and $N_v$ are the normalization constants, $d_{\mu\nu}$ is the polarization sum (the propagator) of the axial-vector diquark, and $L_X^2$ ($X=s$ or $v$) has the form
\begin{align}
L_X^2=(1-x)\Lambda_{X}^2 +x M_{X}^2-x(1-x)M^2,
\end{align}
with $\Lambda_X$ being the cutoff parameters for the quark momentum and $M_X$ being the mass for the diquarks.
In the above calculation, we have adopted the dipolar form factor for the nucleon-quark-diquark couplings.

For calculating the $T$-odd distributions $e_T(x,\bk^2)$ and $e_T^\perp (x,\bk^2)$, one has to consider the nontrivial effect of the gauge link~\cite{Brodsky:2002cx,jy02,Collins:2002plb}, that is, the final-state interaction between the struck quark and the spectator diquark.
Here, we expand the gauge link to one-loop level, as shown in the lower panel of Fig.~\ref{diagrams}.
After some algebra, we arrive at the expressions for the correlator contributed by the scalar and the axial-vector diquark components in the one-gluon exchange approximation, respectively,
\begin{align}
  \Phi_s^{(1)}
(x,\bm k_T)
&\equiv
-i e_q N_{s}^2 {(1-x)^2\over 64\pi^3 (P^+)^2}\frac{-i\Gamma^{+}_s}{(\bm{k}_T^2+L_s^2)^2}\nonumber \\
\hspace{-1cm}&\times \int {d^2 \bm q_T\over (2\pi)^2}
{ \left[(\kslash -q\ssh+m)\gamma_5 S\ssh (\Pslash+M)(\kslash +m)\right]
\over \bm q_{T}^2  \left[(\bm{k}_T-\bm{q}_T)^2+L_s^2\right]^2}
 , \label{phis1}\\
 \Phi^{(1)}_{v}
(x,\bm k_T)
&\equiv
-i e_q N_v^2 \frac{(1-x)^2}{128\pi^3 (P^+)^2}\frac{-i\Gamma^{+,\alpha\beta}}{(\bm{k}_T^2+L_v^2)^2}\nonumber\\
&\times\int {d^2 \bm q_T\over (2\pi)^2} \,
 d_{\rho\alpha}(P-k) d_{\sigma\beta}(P-k+q) \nonumber\\
&\times{ \left[(\kslash -q\ssh+m) \gamma^\sigma \gamma_5 S\ssh(M-\Pslash)\gamma^\rho (\kslash +m)\right]
\over \bm q_T^2  \left[(\bm{k}_T-\bm{q}_T)^2+L_v^2\right]^2},
\label{phia1}
\end{align}
where $q^+=0$, and $e_q$ is the charge for the quarks.
$\Gamma_s^\mu $ or $\Gamma_v^{\mu,\alpha\beta}$ stands for
the vertex between the gluon and the scalar diquark or the axial-vector diquark,
\begin{align}
 \Gamma_s^\mu &= ie_s (2P-2k+q)^\mu, \\
 \Gamma_v^{\mu,\alpha\beta} &=  -i e_v [(2P-2k+q)^\mu g^{\alpha\beta}-(P-k+q)^{\alpha}g^{\mu\beta}\nonumber\\
 &-(P-k)^\beta g^{\mu\alpha}]\label{Gamma},
\end{align}
where $e_{s/v}$ denotes the charge of the scalar/axial-vector diquark.

Substituting (\ref{lophis}) into (\ref{gTs}), we obtain the $T$-even distributions $g_T$ and $g_T^\perp$ from the scalar diquark component,
\begin{align}
g_T^s(x,\bk^2)&=\frac{{N_s}^2(1-x)^2}{16\pi^3}
\frac{1}{(\bk^2+L_s^2)^4}\nonumber\\
&\times\left[(1-x)(m+M)(m+x M)\right.\nonumber\\
&-\left.(x+\frac{m}{M})(\bk^2+M_s^2)\right],\label{eq:gts}\\
g_T^{\perp s}(x,\bk^2)&=\frac{N_s^2(1-x)^3}{8\pi^3}\frac{M^2}{(\bk^2+L_s^2)^4},\label{eq:gtperps}
\end{align}
which are consistent with the results in Ref.~\cite{Jakob:1997wg}.
Similarly, substituting (\ref{lophis}) into (\ref{eTperp}) and (\ref{eT}), we get the $T$-odd distributions $e_T$ and $e_T^\perp$ from the scalar diquark component,
\begin{align}
e_T^s(x,\bk^2)&=-\frac{{N_s}^2(1-x)^2}{32\pi^3} \frac{e_s e_q}{4\pi}\frac{1}{L_s^2(L_s^2+\bk^2)^3}\nonumber\\
&\times\left[(1-x)(2mM+M^2+xM^2)-L_s^2-M_s^2\right],\label{eq:ets}\\
e_T^{\perp s}(x,\bk^2)&=\frac{{N_s}^2(1-x)^2}{32\pi^3} \frac{e_s e_q}{4\pi}\frac{(1-x)^2 M^2-L_s^2-M_s^2}{L_s^2(L_s^2+\bk^2)^3}.\label{eq:etperps}
\end{align}

To calculate the quark correlator contributed by the axial-vector diquark, we choose the form for the propagator $d_{\mu\nu}$ as
\begin{align}
 d_{\mu\nu}(P-k)  =& \,-g_{\mu\nu}\,+\, {(P-k)_\mu n_{-\nu}
 \,+ \,(P-k)_\nu n_{-\mu}\over(P-k)\cdot n_-}\,\nonumber\\
 & - \,{M_v^2 \over\left[(P-k)\cdot n_-\right]^2 }\,n_{-\mu} n_{-\nu} ,\label{d1}
\end{align}
which is the summation over the light-cone transverse polarizations of the axial-vector diquark~\cite{Brodsky:2000ii} and has been applied to calculate leading-twist TMD distributions in Ref.~\cite{Bacchetta:2008af}.
We note that other forms for $d_{\mu\nu}$ have been chosen in Refs.~\cite{Jakob:1997wg,Bacchetta:2003rz,Gamberg:2007wm}.
Using the propagator~(\ref{d1}), we obtain the expressions for the twist-3 TMD distributions from the axial-vector diquark component,
\begin{align}
g_T^v(x,\bk^2)&=\frac{N_v^2(1-x)}{16\pi^3}\frac{\left[x(1-x)-(1+x)\frac{m}{M}\right]\bk^2}{(L_v^2+\bk^2)^4},\label{gTa}\\
g_T^{\perp v}(x,\bk^2)&=\frac{N_v^2(1-x)^2}{8\pi^3}\frac{M(m+xM)}{(L_v^2+\bk^2)^4},
\label{gTperpa}\\
e_T^v(x,\bk^2)&=\frac{N_v^2(1-x)}{32\pi^3 L_v^2(L_v^2+\bk^2)^3}\frac{e_v e_q}{4\pi}\nonumber\\
&\times\left[\frac{(1-x)(m^2-xM^2)-L_v^2+xM_v^2}{L_v^2 (L_v^2+\bk^2)}\right.\nonumber\\
&+\left.\frac{1}{\bk^2}\ln{\frac{\bk^2+L_v^2}{L_v^2}}\right],
\label{eTa}\\
e_T^{\perp v}(x,\bk^2)&=-\frac{N_v^2(1-x)}{32\pi^3(L_v^2+\bk^2)^2}\frac{e_v e_q}{4\pi}\nonumber\\
&\times\left[\frac{(1-x)(m^2+2xmM+xM^2)+L_v^2-xM_v^2}{L_v^2 (L_v^2+\bk^2)}\right.\nonumber\\
&-\left.\frac{1}{\bk^2}\ln{\frac{\bk^2+L_v^2}{L_v^2}}\right].
\label{eTperpa}
\end{align}

\begin{figure}
   \includegraphics[width=0.49\columnwidth]{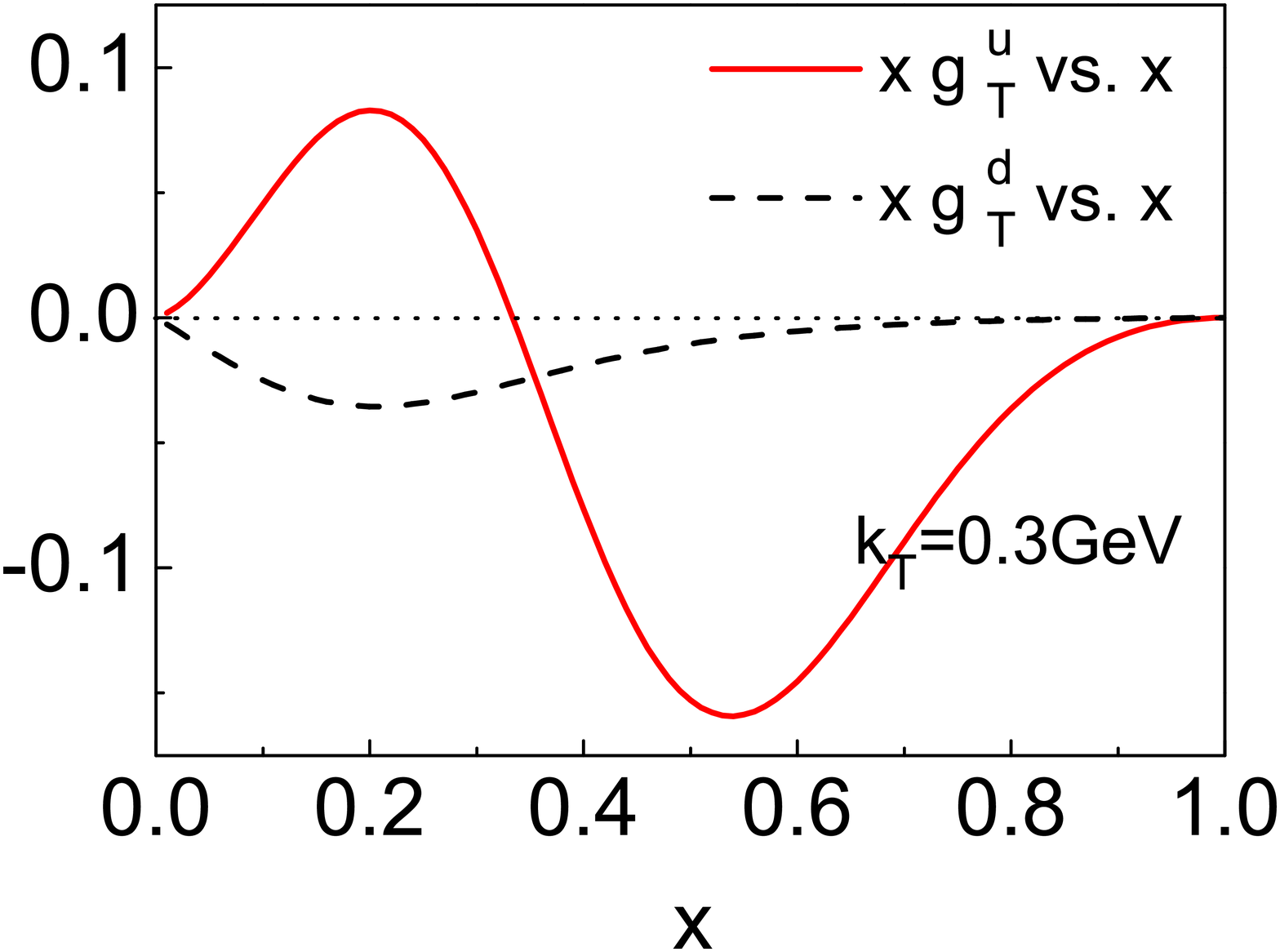}
  \includegraphics[width=0.49\columnwidth]{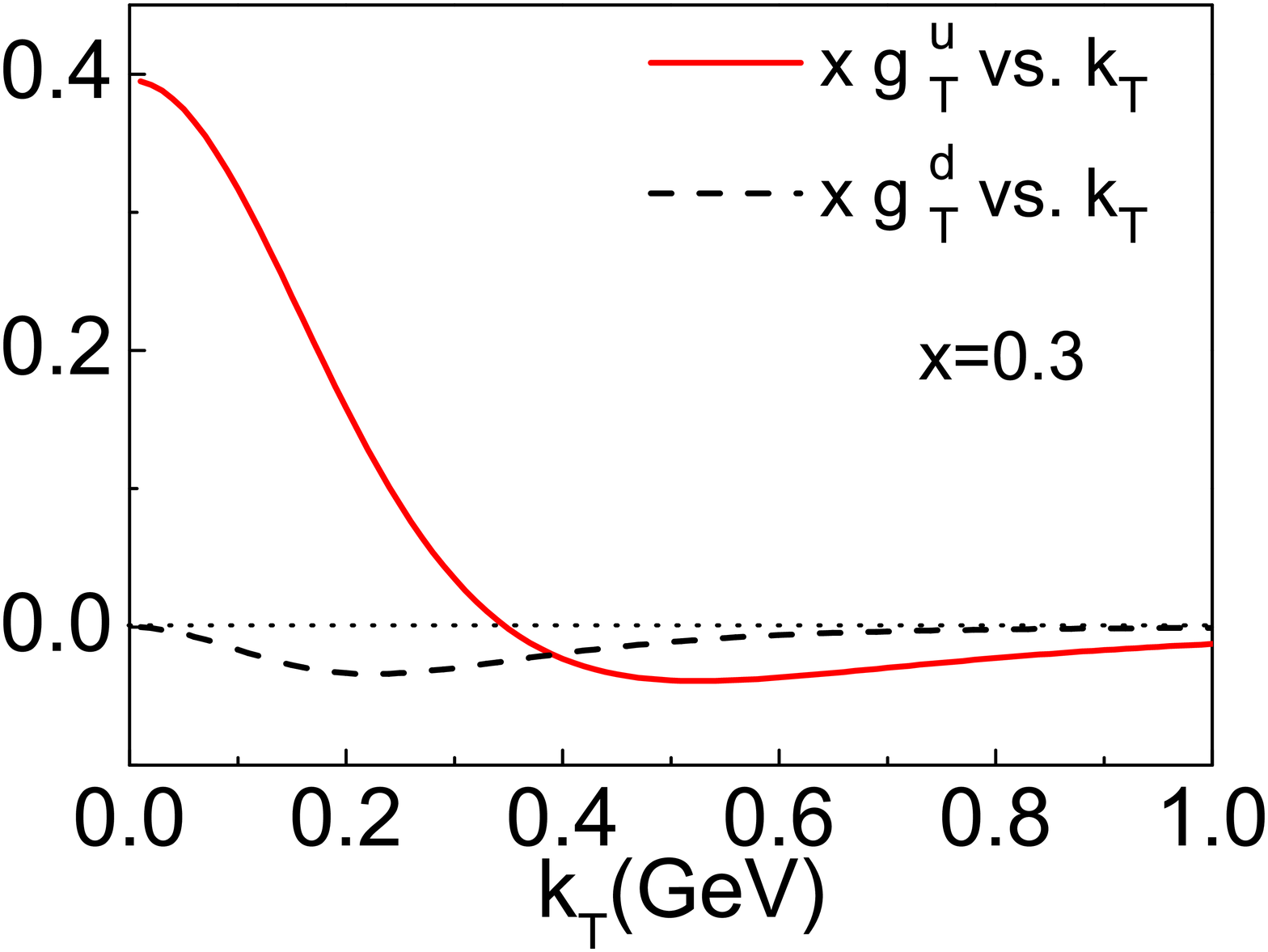}
  \caption{Left panel: Model results for $x g_T^{u}$ (solid line) and $x g_T^{d}$ (dashed line) as functions of $x$ at $k_T=0.3$ GeV; right panel: model results for $x g_T^{u}$ (solid line) and $x g_T^{d}$ (dashed line) as functions of $k_T$ at $x=0.3$.}\label{FIG:gt}
\end{figure}
\begin{figure}
  \includegraphics[width=0.49\columnwidth]{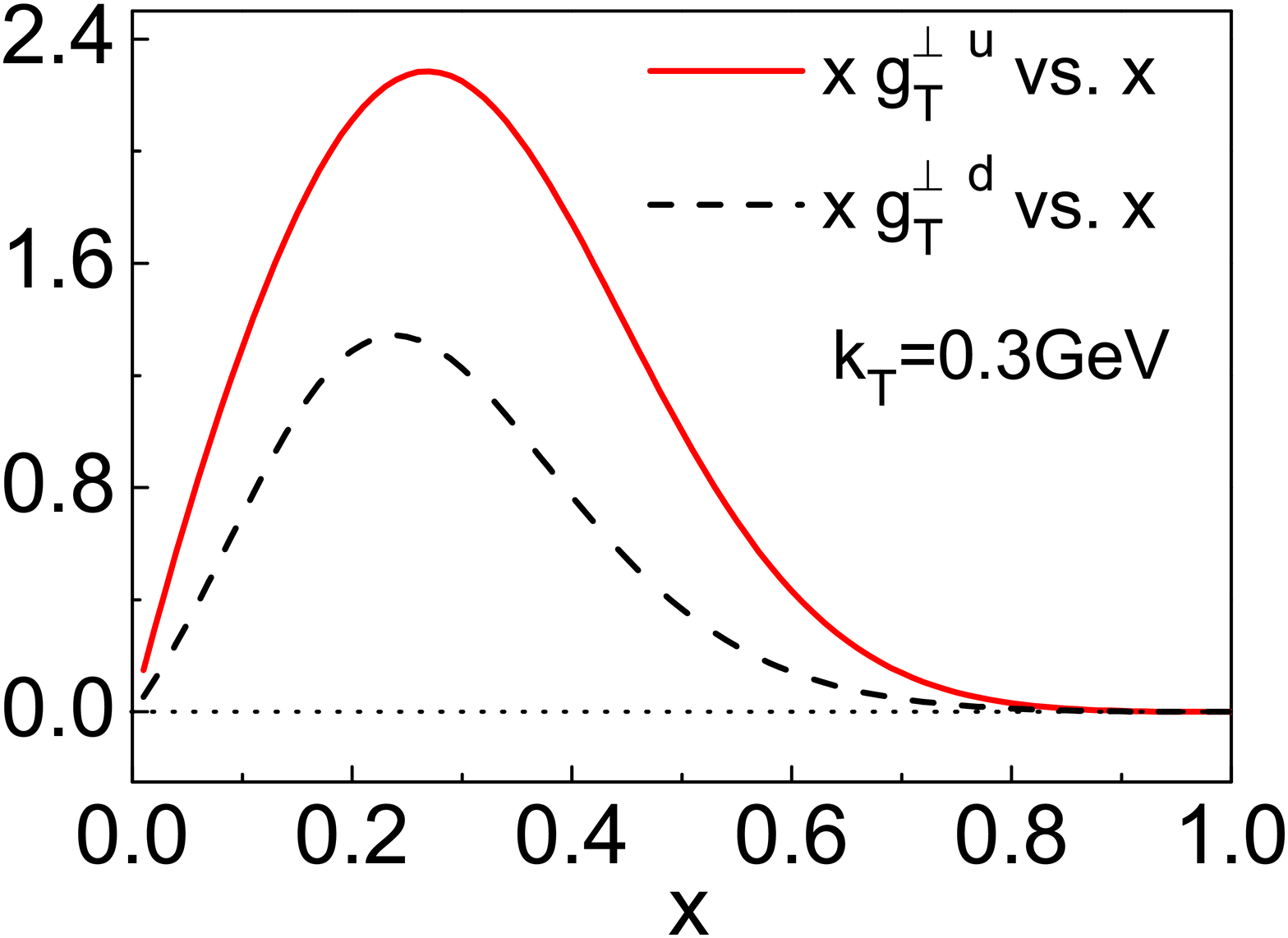}
  \includegraphics[width=0.49\columnwidth]{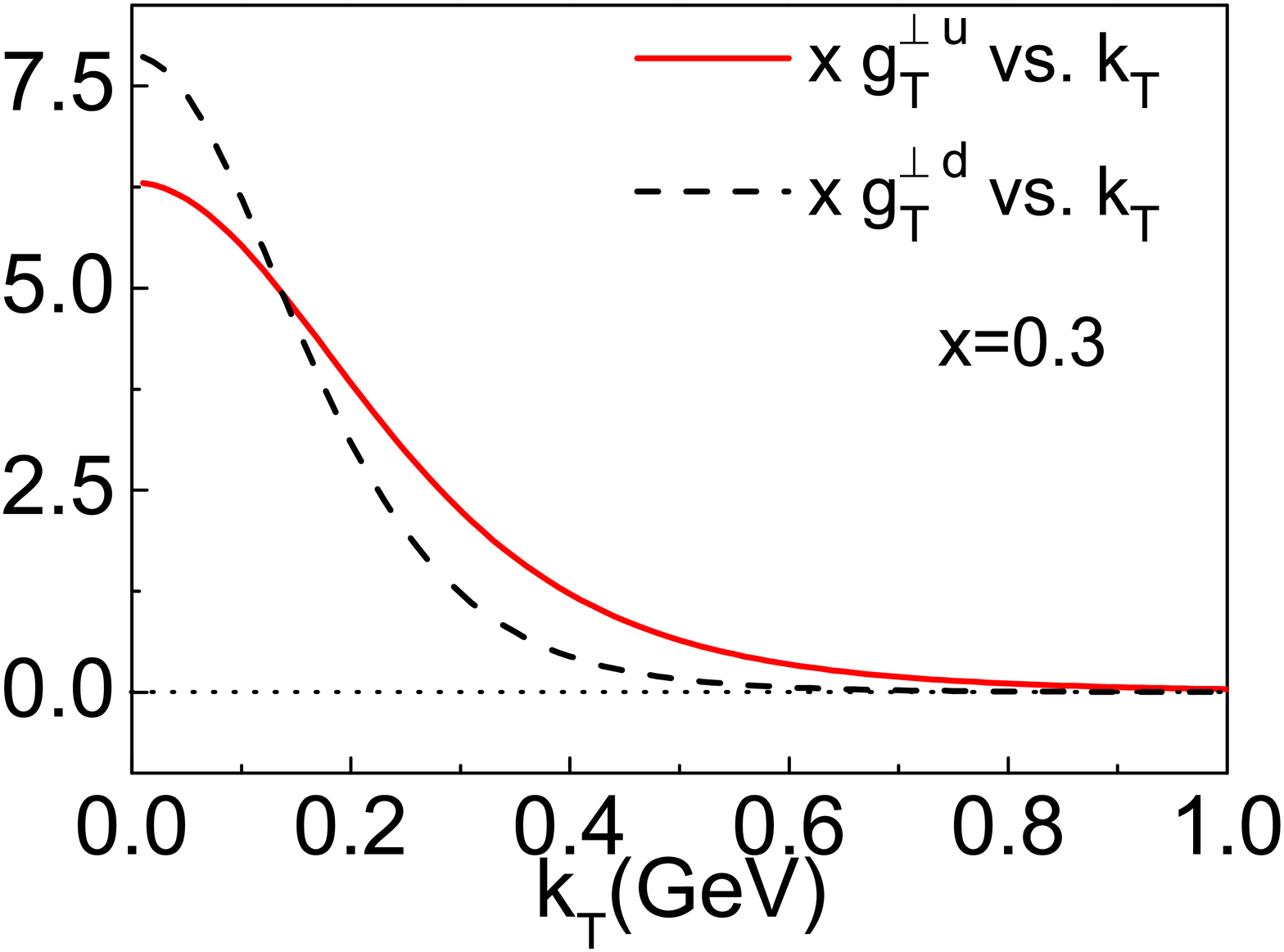}
  \caption{Similar to Fig.~\ref{FIG:gt}, but for the model results of $x g_T^{\perp u}$ (solid line) and $x g_T^{\perp d}$ (dashed line).}\label{FIG:gtperp}
\end{figure}
\begin{figure}
  \includegraphics[width=0.49\columnwidth]{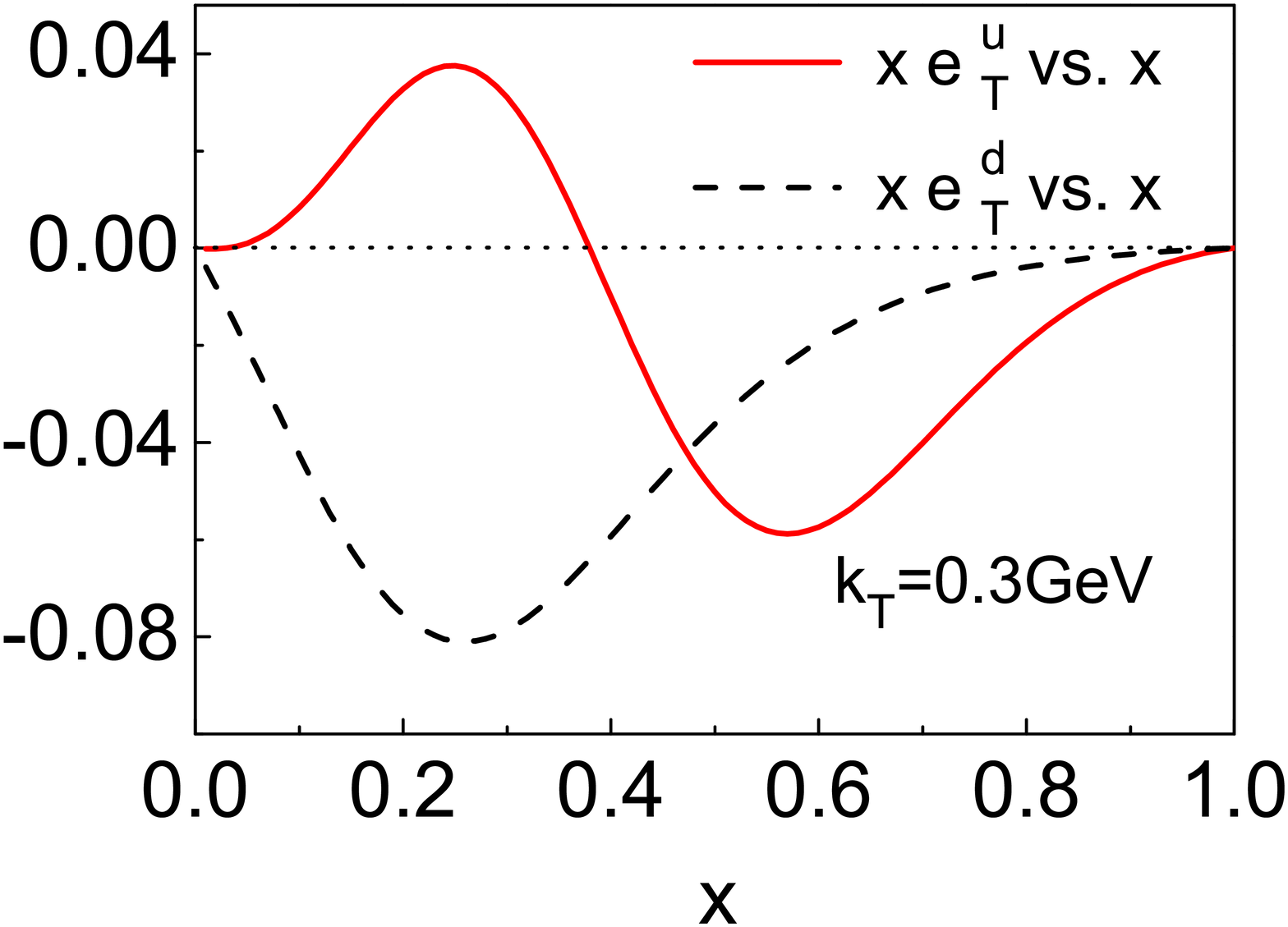}
  \includegraphics[width=0.49\columnwidth]{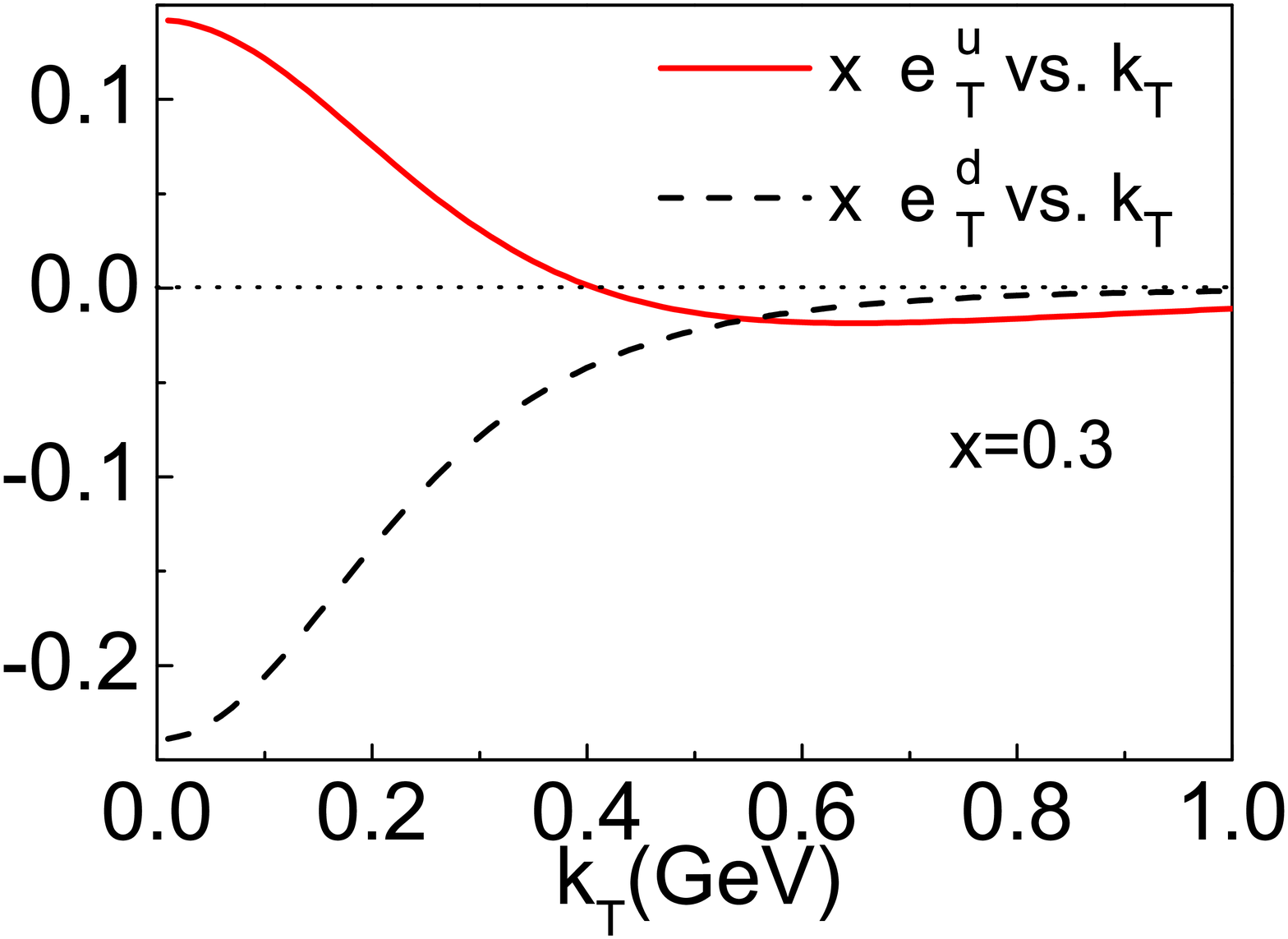}
  \caption{Similar to Fig.~\ref{FIG:gt}, but for the model results of $x e_T^{u}$ (solid line) and $x e_T^{d}$ (dashed line).}\label{FIG:et}
\end{figure}
\begin{figure}
  \includegraphics[width=0.49\columnwidth]{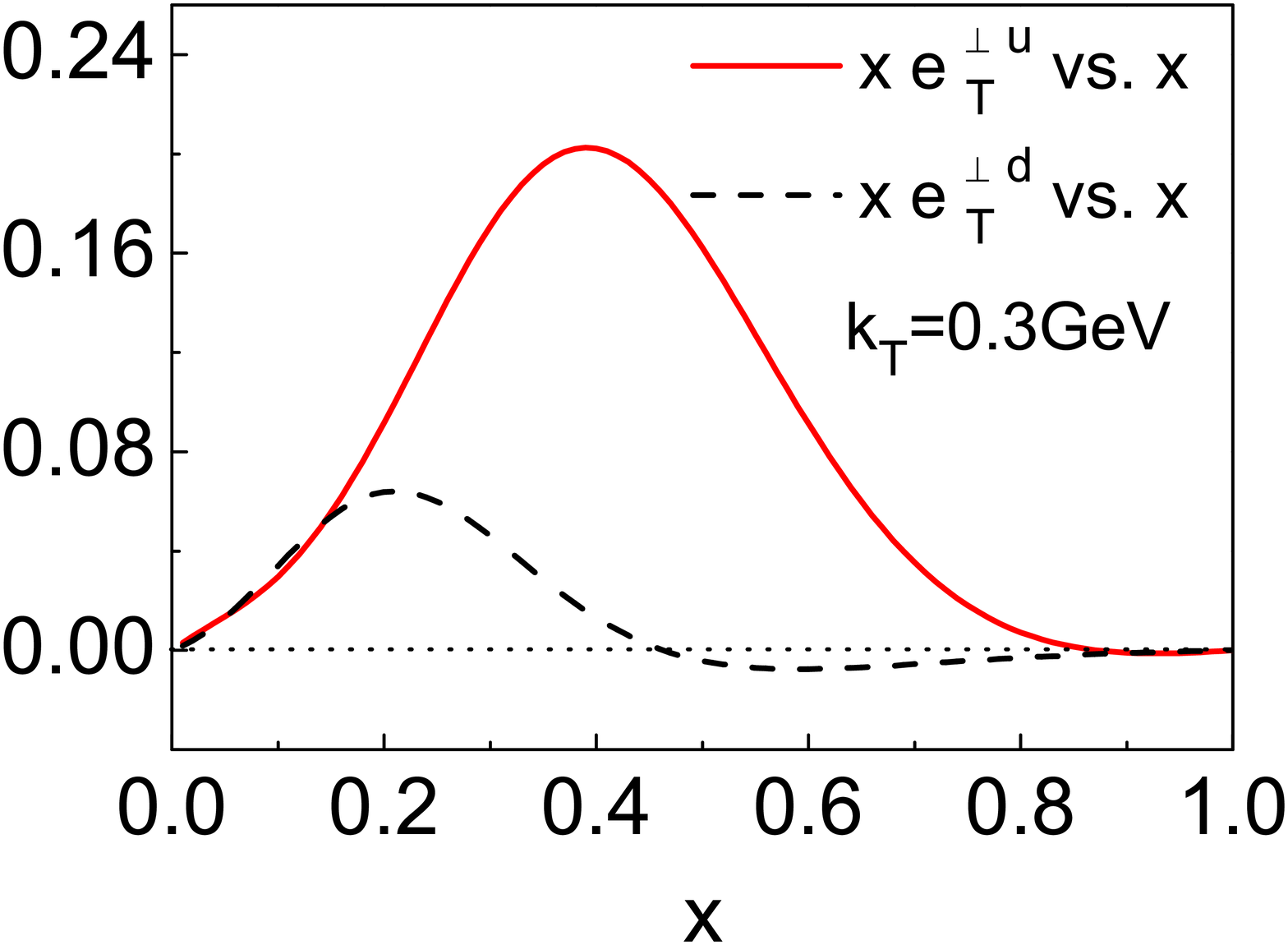}
  \includegraphics[width=0.49\columnwidth]{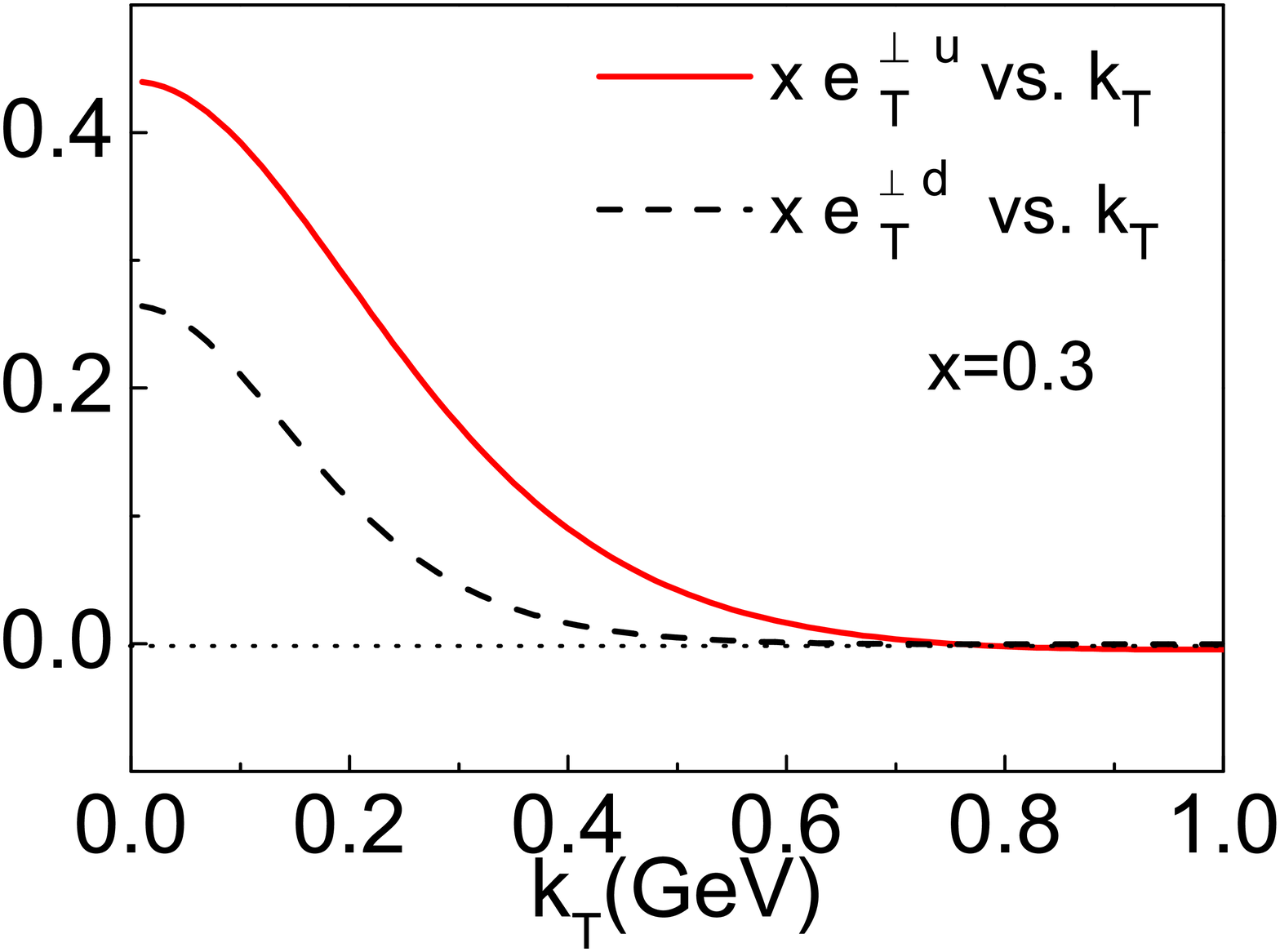}
  \caption{Similar to Fig.~\ref{FIG:gt}, but for the model results of $x e_T^{\perp u}$ (solid line) and $x e_T^{\perp d}$ (dashed line).}\label{FIG:etperp}
\end{figure}
To construct the distributions for the $u$ and $d$ valence quarks from $f^{s}$ and $f^{v}$,
here we follow the approach used in Ref.~\cite{Bacchetta:2008af},
\begin{align}
f^u=c_s^2 f^s + c_a^2 f^a,~~~~f^d=c_{a^\prime}^2 f^{a^\prime}\label{ud},
\end{align}
which gives a general relation between the quark flavor and the diquark type.
Here $a$ and $a^\prime$ denote the vector isoscalar diquark $a(ud)$ and the vector isovector diquark $a(uu)$, respectively, and $c_s$, $c_a$, and $c_{a^\prime}$ are the free parameters of the model and are adopted from Ref.~\cite{Bacchetta:2008af}.
Finally, to convert our calculation to the context of QCD color interaction, we apply the following replacement for the combination of the charges of the quark $q$ and the spectator diquark $X$:
\begin{align}
{e_qe_X\over 4 \pi}\rightarrow -  C_F \alpha_s,
\end{align}
and we choose the coupling constant $\alpha_s \approx 0.3$.

In the left and right panels of Figs.~\ref{FIG:gt},~\ref{FIG:gtperp},~\ref{FIG:et}, and~\ref{FIG:etperp}, we plot the $x$ dependence (at $k_T=0.3$ GeV) and $k_T$ dependence (at $x=0.3$) of the four distributions $g_T$, $g_T^\perp$, $e_T$, and $e_T^\perp$ timed with $x$.
The solid and dashed curves show the results for the $u$ and $d$ valence quarks, respectively.
We find that in the spectator model we applied, generally the sizes of the $T$-even distributions $g_T^\perp$ and $g_T$ are larger than those of the $T$-odd distributions $e_T^\perp$ and $e_T$.
For both the $x$ and $k_T$ dependencies, $g_T^\perp$ and $e_T^\perp$ have the similar shapes and are positive for $u$ and $d$ quarks.
Especially, $g_T^u$ or $e_T^u$ has a node in the $x$-dependent and $k_T$-dependent curves; while $g_T^d$ and $e_T^d$ turn out to be negative.

\section{Prediction on the transverse target DSAs for charged and neutral pions in SIDIS}
\label{BSAs}

\begin{figure}
  \includegraphics[width=0.8\columnwidth]{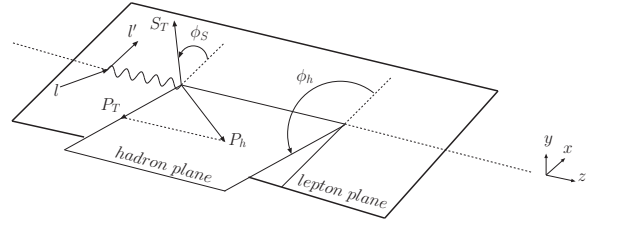}
 \caption {The kinematical configuration for the polarized SIDIS process. The initial and scattered leptonic momenta define the lepton plane ($x-z$ plane), while the detected hadron momentum together with the $z$ axis identify the hadron production plane.}
 \label{SIDISframe}
\end{figure}

In this section, we will show our predictions on the transverse target DSAs in SIDIS, and we limit our attention on the asymmetries of pion production at the subleading-twist level.
The process of scattering a longitudinal polarized lepton beam off a transversely polarized target can be expressed as
\begin{align}
l^\rightarrow (\ell) \, + \, p^\uparrow (P) \, \rightarrow \, l' (\ell')
\, + \, h (P_h) \, + \, X (P_X)\,,
\label{sidis}
\end{align}
where $\rightarrow$ represents the longitudinal polarization of the lepton beam, and $\uparrow$ represents the transverse polarization of the proton target; $\ell$ and $\ell'$ denote the momenta of the incoming and outgoing leptons, and $P$ and $P_h$ represent the momenta of the target nucleon and the final-state hadron.
In our calculation, we adopt the reference frame shown in Fig.~\ref{SIDISframe}, following the Trento convention~\cite{Bacchetta:2004jz}.
Within this reference frame, $\bm P_T$ and $\bm {S}_T$ denote the transverse momenta of the detected pion the transverse spin of the target, and $\phi_h$ and $\phi_S$ denote their azimuthal angles with respect to the lepton plane, respectively.

The invariant variables used to express the kinematics of SIDIS under study are defined as
\begin{align}
&x = \frac{Q^2}{2\,P\cdot q},~~~
y = \frac{P \cdot q}{P \cdot l},~~~
z = \frac{P \cdot P_h}{P\cdot q},~~~
\gamma={2M x\over Q},~~~\nonumber\\
&Q^2=-q^2, ~~~
s=(P+\ell)^2,~~~
W^2=(P+q)^2,~~~
\end{align}
where $q=\ell-\ell'$ is the four momentum of the virtual photon, and $W$ is the invariant mass of the hadronic final state.
The differential cross section of the process (\ref{sidis}) can be expressed as~\cite{Bacchetta:2006tn}
\begin{align}
\frac{d\sigma}{d\xbj dy\,d{\phi_S}dz dP^2_T d\ph} &=\frac{\alpha^2}{\xbj y Q^2}\frac{y^2}{2(1-\varepsilon)}
 \Bigl( 1+ \frac{\gamma^2}{2\xbj} \Bigr)
  \left\{F_{\mathrm{UU}} \right.\nonumber\\
  &+ \left.|\bm{S}_T| \left[\sqrt{2\,\varepsilon (1-\varepsilon)}\,\left(\cos\phi_S\,F_{\mathrm{LT}}^{\cos \phi_S }\right.\right.\right.\nonumber\\
&+ \left.\left.
 \left. \cos(2\phi_h-\phi_S)\,
F_{\mathrm{LT}}^{\cos{(2\phi_h -\phi_S)}}
\right)\right]\right.\nonumber\\
&+\left.\cdots\right\}.\label{FLT}
\end{align}
Here, $F_{\mathrm{UU}}$ is the spin-averaged structure function, and $F_{\mathrm{LT}}^{\cos{\phi_S}}$ and $F_{\mathrm{LT}}^{\cos{(2\phi_h -\phi_S)}}$ are the spin-dependent structure functions that contribute to the $\cos{\phi_S}$ and $\cos{(2\phi_h-\phi_s)}$ azimuthal asymmetries, respectively. The ellipsis stands for the leading-twist contribution to the $\cos{(\phi_h - \phi_S)}$ moment, which is contributed by the $g_{1T}^\perp D_1$ term and will not be studied in this work.
The ratio of the longitudinal and transverse photon flux $\varepsilon$ is defined as
\begin{align}
\varepsilon=\frac{1-y-\gamma^2y^2/4}{1-y+y^2/2+\gamma^2y^2/4}.
\end{align}

In the parton model, the structure functions in Eq.~(\ref{FLT}) can be expressed as the convolutions of twist-2 and twist-3 TMD distributions and FFs (based on the tree-level factorization from Ref.~\cite{Bacchetta:2006tn}).
With the adopted reference frame and the notation
\begin{align}
\mathcal{C}[w fD] &=x\sum_q e_q^2\int d^2\bm k_T\int d^2 \bm p_T\delta^2(z\bm k_T-\bm P_T+\bm p_T) \nonumber\\
&\times w(\bm k_T, \bm p_T)f^q(x,\bm k_T^2) D^q(z,\bm p_T^2),
\end{align}
$F_{\mathrm{UU}}$, $F_\mathrm{{LT}}^{\cos{\phi_S}}$, and $F_{\mathrm{LT}}^{\cos{(2\phi_h -\phi_S)}}$ can be expressed in the following forms~\cite{Bacchetta:2006tn}:
\begin{align}
F_{\mathrm{UU}} &= \mathcal{C}[f_1 D_1], \label{FUU}\\
F_{\mathrm{LT}}^{\cos\phi_S}&\approx\frac{2M}{Q}{\cal C}\left\{-x g_T D_1\right.\nonumber\\
&-\left.\frac{\bp \cdot \bk}{2z M M_h}\left(x e_T H_1^\perp+x e_T^\perp H_1^\perp\right)
\right\},
\label{FLT1} \\
F_{\mathrm{LT}}^{\cos{(2\phi_h-\phi_S)}}&\approx\frac{2M}{Q}{\cal C}\left\{-\frac{2(\hat{\bm P}_T\cdot \bk)^2-\bk^2}{2M^2}\left(x g_T^{\perp}D_1\right)\right.\nonumber\\
&-\frac{2(\hat{\bm P}_T \cdot \bp)(\hat{\bm P}_T\cdot\bk)-\bp\cdot \bk}{2zM M_h}\nonumber\\
&\times\left.\left[x e_T H_1^\perp-x e_T^\perp H_1^\perp\right]\right\}.
\label{FLT2}
\end{align}
Here we introduced the unit vector $\hat {\bm P}_T = \frac{\bP}{|\bP|}$ and denoted the mass of the final-state hadron by $M_h$.
We also neglected the contributions from the twist-3 FFs $\tilde{E}$, $\tilde{D}^\perp$, and $\tilde{G}^\perp$ in Eqs.~(\ref{FLT1}) and (\ref{FLT2}), by applying the Wandzura-Wilczek approximation~\cite{Wandzura:1977qf}.
That is, we assume that the contributions from the terms including a twist-3 FF with a tilde are very small.
Therefore, we restrict the scope on the contributions only from the terms containing a twist-3 distribution function in our calculation.
With Eqs.~(\ref{FUU}), (\ref{FLT1}), and (\ref{FLT2}), the $P_T$-dependent  DSAs $A_{\mathrm{LT}}^{\cos\phi_S}$ and $A_{\mathrm{LT}}^{\cos{(2\phi_h-\phi_S)}}$ can be given as
\begin{align}
A_{\mathrm{LT}}^{\cos\phi_S}(P_T) &= \frac{\int dx \int dy \int dz \;\mathcal{C}_{\mathrm{LT}}\;F_{\mathrm{LT}}^{\cos{\phi_S}}}{\int dx \int dy \int dz \;\mathcal{C}_{\mathrm{UU}}\;F_{\mathrm{UU}} }\;,\label{asy1}
\end{align}
\begin{align}
A_{\mathrm{LT}}^{\cos{(2\phi_h-\phi_S)}}(P_T) &= \frac{\int dx \int dy \int dz \;\mathcal{C}_{\mathrm{LT}} \;F_{\mathrm{LT}}^{\cos{(2\phi_h-\phi_S)}}}{\int dx \int dy \int dz \;\mathcal{C}_{\mathrm{UU}}\;F_{\mathrm{UU}} }\;, \label{asy2}
\end{align}
where we have defined the kinematical factors
\begin{align}
\mathcal{C}_{\mathrm{UU}}&=\frac{1}{x y Q^2}\frac{y^2}{2(1-\varepsilon)}\Bigl( 1+ \frac{\gamma^2}{2x}
\Bigl),\\
\mathcal{C}_{\mathrm{LT}}&=\frac{1}{x y Q^2}\frac{y^2}{2(1-\varepsilon)}\Bigl( 1+ \frac{\gamma^2}{2x} \Bigr) \sqrt{2\varepsilon(1-\varepsilon)}.
\end{align}
The $x$-dependent and the $z$-dependent asymmetries can be defined in a similar way.
We need to point out that we have assumed that the TMD factorization can be generalized to twist-3 level to obtain Eqs.~(\ref{FLT1}) and (\ref{FLT2}).
However, from a theoretical point of view, one should keep in mind that it is still not very clear if the TMD factorization is valid when dealing with the higher-twist observables under the TMD framework~\cite{Gamberg:2006ru,Bacchetta:2008xw}.
Nevertheless, since there is no a better way to deal with this problem or an alternative theoretical approach for the transverse target DSAs in the low $P_T$ region, as a first attempt, we would like to use the tree-level results in Ref.~\cite{Bacchetta:2006tn} as a more phenomenological approach to perform the estimates.

To give the numerical predictions on the transverse target DSAs at subleading twist, besides the twist-3 TMD distributions, we also need to know the unpolarized TMD distribution $f_1(x,\bm k_T^2)$, the TMD twist-2 FF $D_1^q(z,\bm p_T^2)$, and the Collins function $H_1^\perp(z,\bm p_T^2)$.
For consistency, we adopt the same model result~\cite{Bacchetta:2008af} for $f_1$, which is fitted from the ZEUS~\cite{Chekanov:2002pv} data set on the unpolarized distribution.
For the TMD FF $D_1^q(z,\bp^2)$, we assume its $p_T$ dependence has a Gaussian form
\begin{align}
D_1^q\left(z,\bp^2\right)=D_1^q(z)\, \frac{1}{\pi \langle \bp^2\rangle}
\, e^{-\bm p_T^2/\langle \bp^2\rangle},
\end{align}
and choose the Gaussian width for $\bp^2$ as $\langle \bp^2\rangle=0.2~\textrm{GeV}^2$, following the result obtained in Ref.~\cite{Anselmino:2005nn}.
For the integrated FF $D_1^q(z)$, we apply the leading-order set of the DSS parametrization~\cite{deFlorian:2007aj}.
As for the Collins functions for different pion productions, we use the following relations:
\begin{align}
 H_1^{\perp \pi^+/u}&=H_1^{\perp \pi^-/d}\equiv H_{1 \mathrm{fav}}^{\perp} ,\\
 H_1^{\perp \pi^+/d}&=H_1^{\perp \pi^-/u}\equiv H_{1 \mathrm{unf}}^{\perp} ,\\
 H_1^{\perp \pi^0/u}&=H_1^{\perp \pi^0/d}\equiv{1\over 2}\left( H_{1 \mathrm{fav}}^{\perp}+H_{1 \mathrm{unf}}^{\perp}\right),
\label{collins}
\end{align}
where $H_{1 \mathrm{fav}}^{\perp}$ and $H_{1 \mathrm{unf}}^{\perp}$ are the favored and unfavored Collins functions, for which we employ the parametrization from Ref.~\cite{Anselmino:2008jk}.

Finally, we also take into consideration the following kinematical constraints~\cite{Boglione:2011wm} on the intrinsic transverse momenta of the initial quarks  throughout our numerical calculation:
\begin{equation}
 \begin{cases}
\bm k_{T}^2\leq(2-x)(1-x)Q^2, ~~~\textrm{for}~~0< x< 1 
; \\
\bm k_{T}^2\leq \frac{x(1-x)} {(1-2x)^2}\, Q^2, ~~~~~~~~~~~~\textrm{for}~~x< 0.5.
\end{cases}\label{constraints}
 \end{equation}
The first constraint is obtained by requiring the energy of the parton to be less than the energy of the parent hadron; the second is given by the requirement that the parton should move in the forward direction with respect to the parent hadron.

\subsection{HERMES}
The kinematical cuts we adopt to perform numerical calculation on the transverse target DSAs at HERMES are as follows~\cite{Airapetian:2009ae}:
\begin{align}
&0.023 < x < 0.4,~~0.1 < y < 0.95,~~0.2 < z < 0.7, \nonumber\\
&W^2 > 10\, \textrm{GeV}^2,~~~Q^2 > 1 \textrm{GeV}^2, \nonumber\\
&0.05 < P_T < 1.2\,\textrm{GeV},~~2\,\textrm{GeV} < E_h < 15\, \textrm{GeV},
\end{align}
where $E_h$ is the energy of the detected pion in the target rest frame.

\begin{figure}
  \includegraphics[width=0.99\columnwidth]{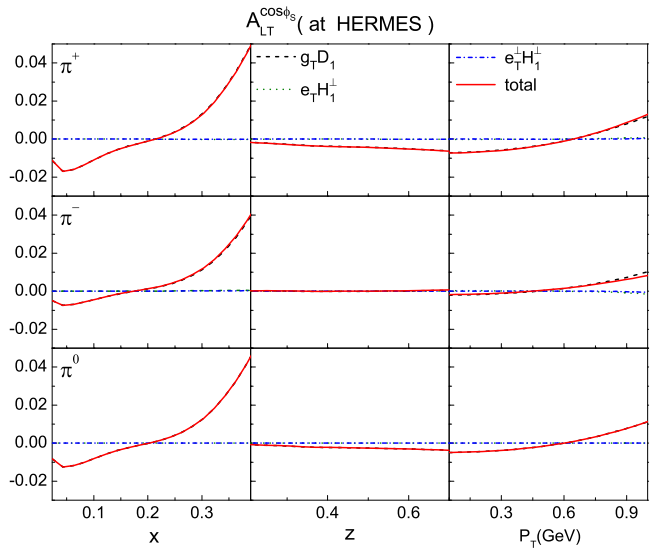}
  \caption{Prediction on the transverse target DSA $A_{\mathrm{LT}}^{\cos\phi_S}$ for $\pi^+$ (upper panel), $\pi^-$ (middle panel), and $\pi^0$ (lower panel) in SIDIS at HERMES. The dashed, dotted, and dash-dotted curves represent the asymmetries from the $g_T D_1$, $e_T H_1^\perp$, and $e_T^\perp H_1^\perp$ terms, respectively. The solid curves correspond to the total contribution.}
  \label{HERMESa}
\end{figure}

In the left, central, and right panels of Fig.~\ref{HERMESa}, we present our estimates of the $\cos \phi_S$ DSA at HERMES for $\pi^+$, $\pi^-$, and $\pi^0$ as functions of $x$, $z$, and $P_T$, respectively.
The dashed, dotted, and dash-dotted curves are used to distinguish the origins from the $g_T D_1$ term, the $e_T H_1^\perp$ term, and the $e_T^\perp H_1^\perp$ term for $A_{\mathrm{LT}}^{\cos \phi_S}$.
The solid curves stand for the total contribution.
As we can see from Fig.~\ref{HERMESa}, the predicted asymmetry is negative in the small $x$ and $P_T$ regions, but turns out to be positive with increasing $x$ and $P_T$,  showing that the asymmetry may be observed in the higher $x$ region.
The $z$-dependent asymmetry is small, which is due to the cancellation of the asymmetry at small and large $x$ ($P_T$).
Moreover, it is the $g_T D_1$ term that gives the dominant contribution to the asymmetry $A_{\mathrm{LT}}^{\cos \phi_S}$, while the contributions from the $e_T H_1^\perp$ term and the $e_T^\perp H_1^\perp$ term are nearly consistent with zero.
This tendency can be found in the overall kinematical regions and for all three pions.
It may be explained by the kinematical factor $\bp \cdot \bk/(2z M M_h)$ associated with the chiral-odd terms, and by the fact that here the sizes of the $T$-even distribution/fragmentation functions are larger than those of the T-odd  distribution/fragmentation functions.

\begin{figure}
  \includegraphics[width=0.99\columnwidth]{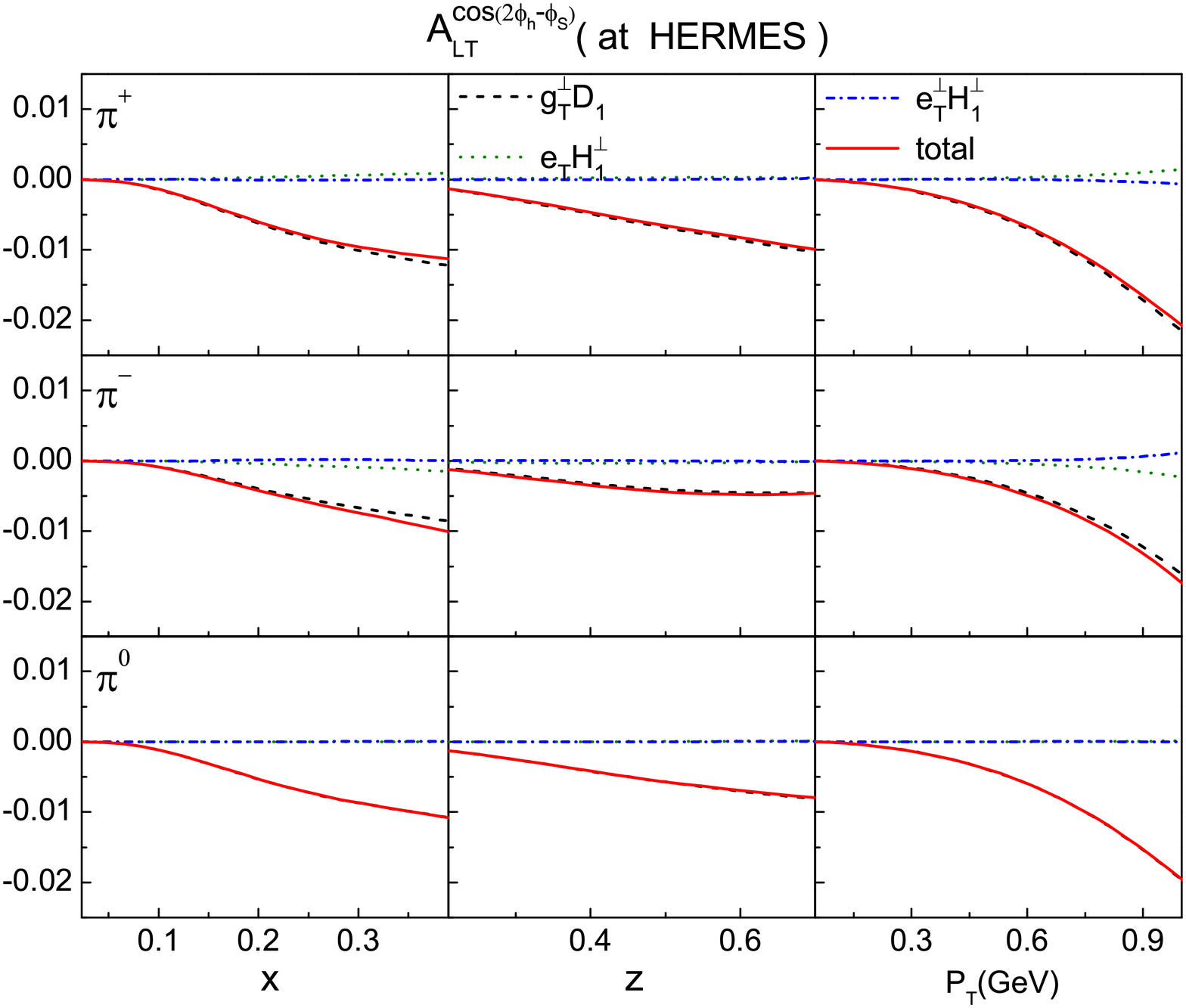}
  \caption{Similar to Fig.~\ref{HERMESa}, but on the asymmetry $A_{\mathrm{LT}}^{\cos(2\phi_h -\phi_S)}$.
  The dashed, dotted, and dash-dotted curves show the asymmetries from the $g_T^\perp D_1$, $e_T H_1^\perp$, and $e_T^\perp H_1^\perp$ terms, respectively. The solid curves correspond to the total contribution.}
  \label{HERMESb}
\end{figure}
In Fig.~\ref{HERMESb}, we plot the asymmetry $A_{\mathrm{LT}}^{\cos(2\phi_h -\phi_S)}$ for $\pi^+$, $\pi^-$ and $\pi^0$ productions.
The contributions from the $g_T^\perp D_1$ term, the $e_T H_1^\perp$ term, and the $e_T^\perp H_1^\perp$ term are denoted by the dashed, dotted, and dash-dotted curves, respectively.
We find that the asymmetries are negative, with the sizes around $1\%$ to $2\%$, and increase with increasing $x$, $z$, and $P_T$ in the kinematical region of HERMES.
Similar to the case of the $\cos \phi_S$ asymmetry, the main contribution to $A_{\mathrm{LT}}^{\cos(2\phi_h -\phi_S)}$ for charged and neutral pions is from the $T$-even distribution $g_T$ combined with $D_1$, and the contributions from the $e_T H_1^\perp$ term and the $e_T^\perp H_1^\perp$ term are nearly negligible, except for the asymmetries for charged pions in the larger $P_T$ region.

\subsection{JLab $5.5$ GeV}

\begin{figure}
\includegraphics[width=0.99\columnwidth]{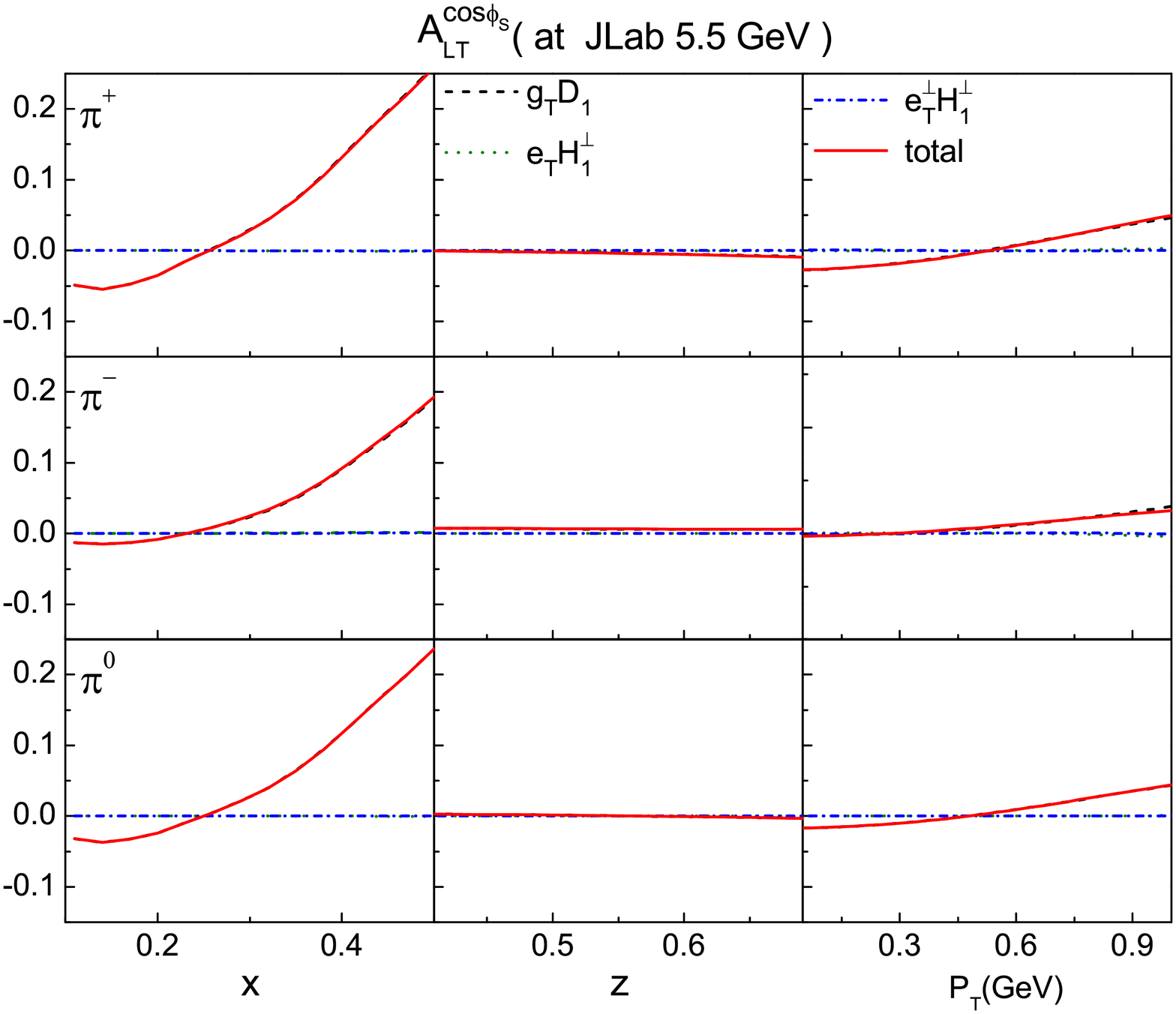}
\caption{Prediction on the transverse target DSA $A_{\mathrm{LT}}^{\cos\phi_S}$ for $\pi^+$ (upper panel), $\pi^-$ (middle panel), and $\pi^0$ (lower panel) in SIDIS at JLab 5.5 GeV.
The dashed, dotted, and dash-dotted curves represent the asymmetries from the $g_T D_1$, $e_T H_1^\perp$, and $e_T^\perp H_1^\perp$ terms, respectively.
The solid curves correspond to the total contribution.}
\label{JLab55a}
\end{figure}
\begin{figure}
\includegraphics[width=0.99\columnwidth]{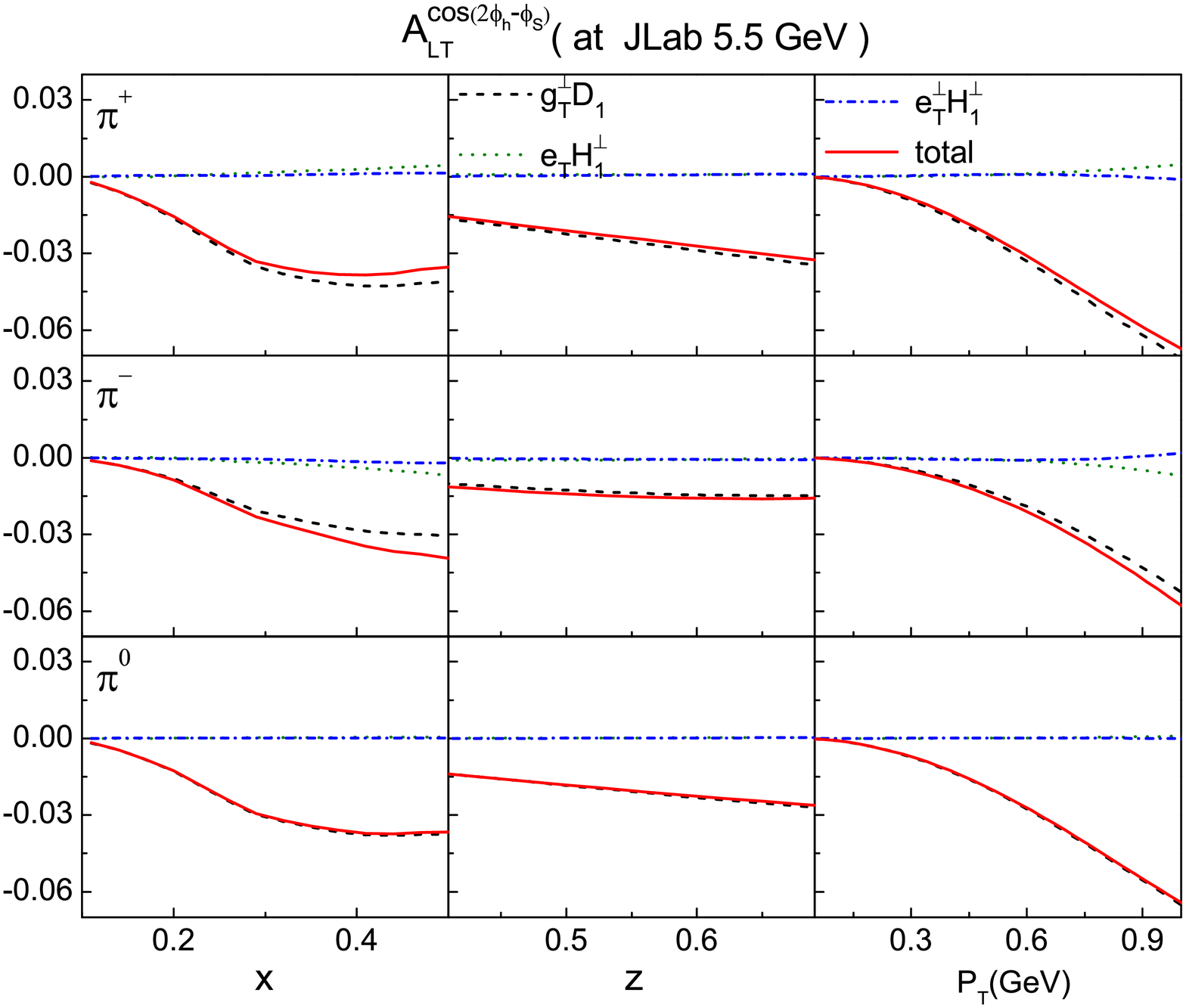}
\caption{Similar to Fig.~\ref{JLab55a}, but on the asymmetry $A_{\mathrm{LT}}^{\cos(2\phi_h -\phi_S)}$.
The dashed, dotted, and dash-dotted curves represent the asymmetries from the $g_T^\perp D_1$, $e_T H_1^\perp$, and $e_T^\perp H_1^\perp$ terms, respectively.
The solid curves correspond to the total contribution.}
\label{JLab55b}
\end{figure}

To test the feasibility to measure the transverse-target DSAs $A_{\mathrm{LT}}^{\cos \phi_S}$
and $A_{\mathrm{LT}}^{\cos(2\phi_h -\phi_S)}$, we also estimate these two asymmetries at the kinematics available at JLab with a $5.5$-GeV electron beam.
The following cuts are the kinematics we adopt in the calculation~\cite{Avakian:2013sta}:
\begin{align}
&0.1<x<0.6,~~ 0.4<z<0.7,~~ Q^2>1\, \textrm{GeV}^2,\nonumber\\
&P_T>0.05\,\textrm{GeV},~~ W^2>4\,\textrm{GeV}^2.
\end{align}

In Figs.~\ref{JLab55a} and \ref{JLab55b}, we plot our estimates on $A_{\mathrm{LT}}^{\cos\phi_S}$ and $A_{\mathrm{LT}}^{\cos(2\phi_h -\phi_S)}$ at JLab for $\pi^+$, $\pi^-$, and $\pi^0$ as functions of $x$, $z$, and $P_T$, respectively.
We find that the magnitude of the asymmetry $A_{\mathrm{LT}}^{\cos\phi_S}$ in Fig.~\ref{JLab55a} as a function of $x$ is large at JLab, around $10\%$, but its $z$ and $P_T$ dependencies are not so obvious.
Again we find that the contribution from the $g_T D_1$ term dominates over the ones from the $e_T H_1^\perp$ term and the $e_T^\perp H_1^\perp$ term.

As for the $A_{\mathrm{LT}}^{\cos(2\phi_h -\phi_S)}$, the results show that sizable asymmetry for pions may be observed at JLab.
In our calculation, the origin of this asymmetry is mainly from the $g_T^\perp D_1$ term, although the $e_T H_1^\perp$ and $e_T^\perp H_1^\perp$ terms give small contributions to $A_{\mathrm{LT}}^{\cos(2\phi_h -\phi_S)}$ at higher $P_T$ for charged pion production.
The asymmetries for all three pions are negative, and the magnitudes appear more sizable as the kinematical variables increase.

\subsection{COMPASS}
\begin{figure}
  \includegraphics[width=0.99\columnwidth]{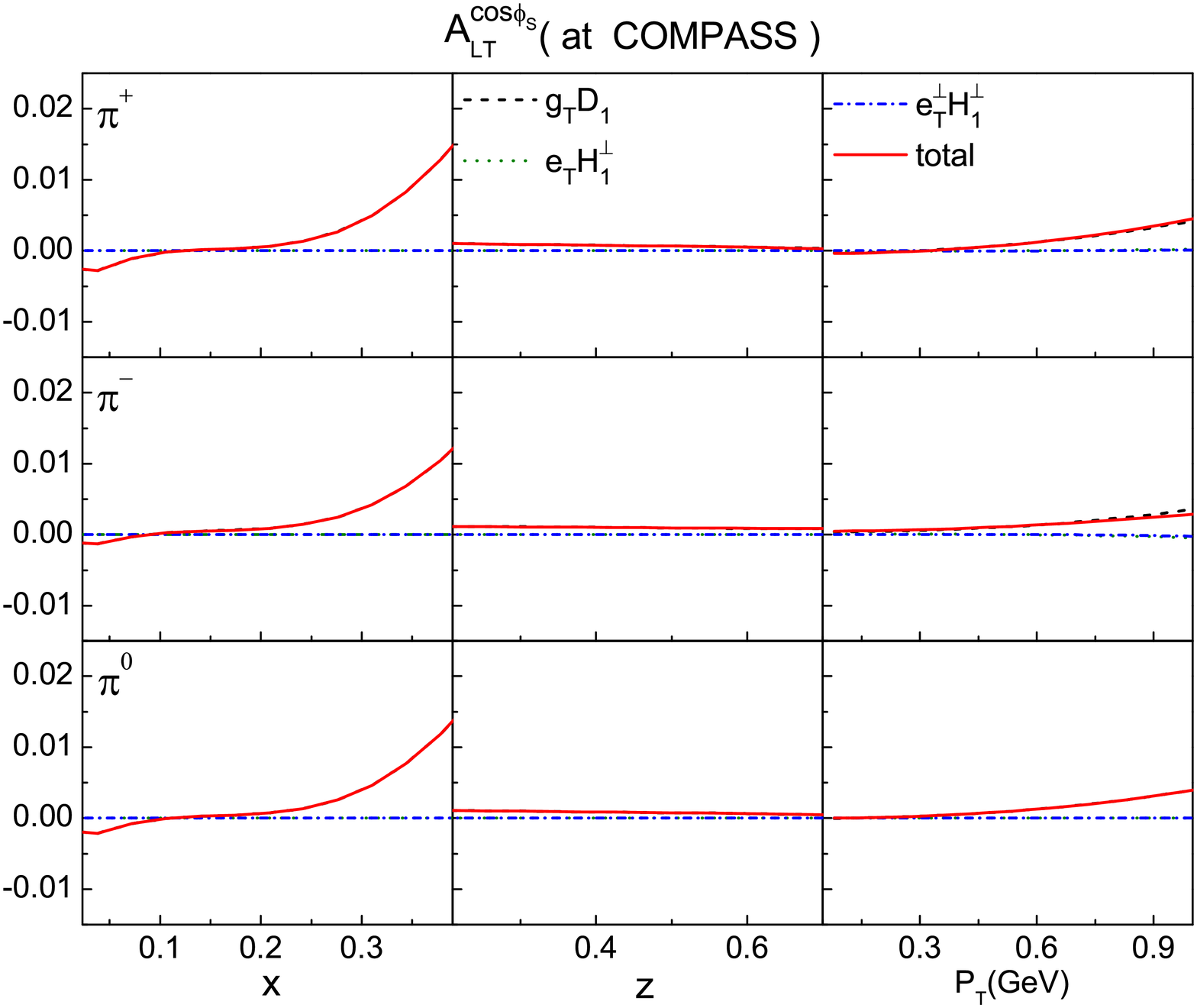}
  \caption{Predictions on the transverse target DSA $A_{\mathrm{LT}}^{\cos{\phi_S}}$ for $\pi^+$ (upper panel), $\pi^-$ (middle panel), and $\pi^0$ (lower panel) in SIDIS at COMPASS.
  The dashed, dotted, and dash-dotted curves represent the asymmetries from the $g_T D_1$, $e_T H_1^\perp$, and $e_T^\perp H_1^\perp$ terms, respectively.
  The solid curves correspond to the total contribution.}
  \label{compassa}
\end{figure}

\begin{figure}
  \includegraphics[width=0.99\columnwidth]{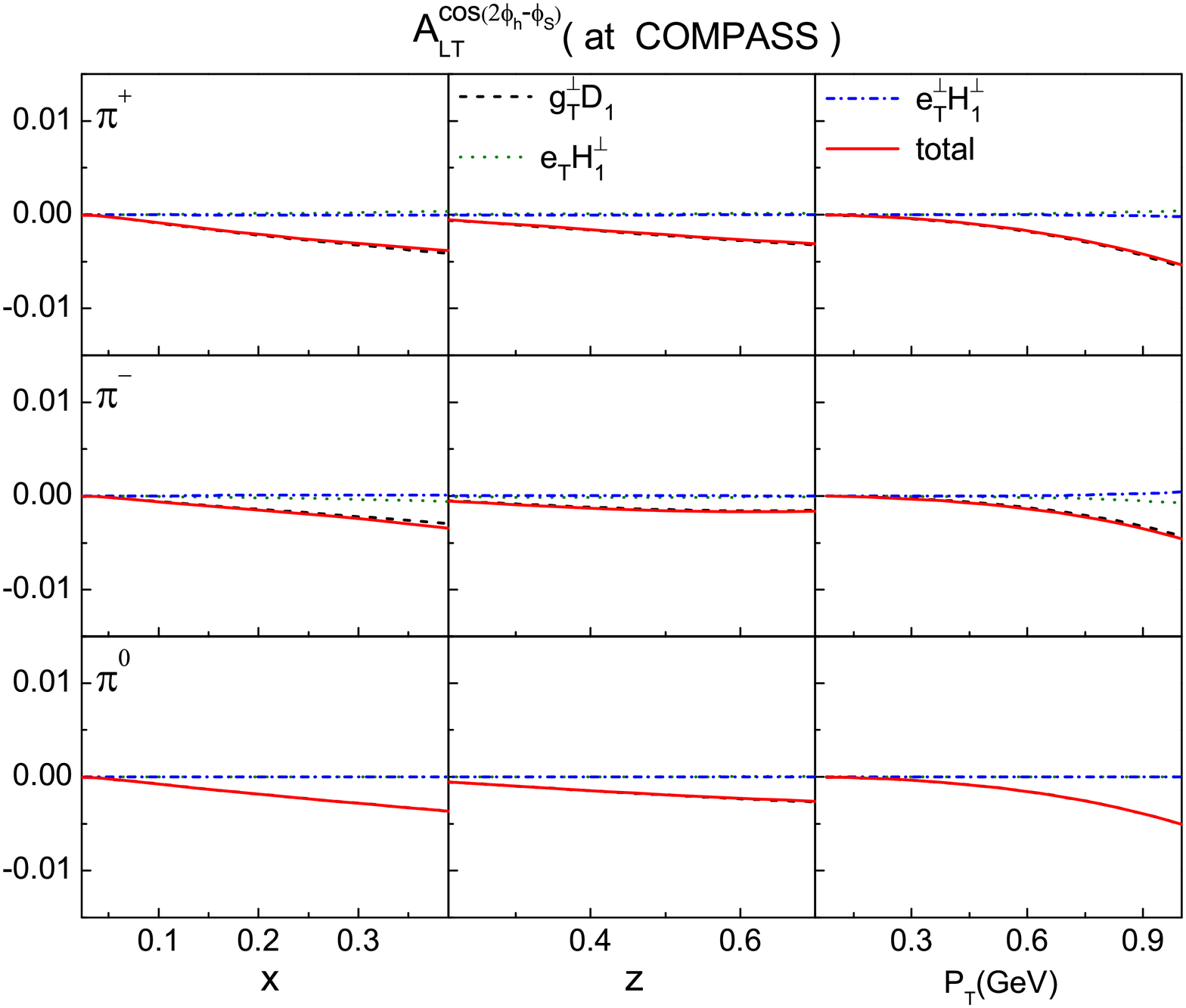}
  \caption{Similar to Fig.~\ref{compassa}, but on the asymmetry $A_{\mathrm{LT}}^{\cos{(2\phi_h -\phi_S)}}$.
  The dashed, dotted, and dash-dotted curves represent the asymmetries from the $g_T^\perp D_1$, $e_T H_1^\perp$, and $e_T^\perp H_1^\perp$ terms, respectively.
  The solid curves correspond to the total contribution.}
  \label{compassb}
\end{figure}

We also make the prediction on the same asymmetries at COMPASS, with a muon beam of 160 GeV scattered off a proton target.
The results for the asymmetries $A_{\mathrm{LT}}^{\cos\phi_S}$ and $A_{\mathrm{LT}}^{\cos(2\phi_h -\phi_S)}$ are shown in Figs.~\ref{compassa} and~\ref{compassb}, respectively.
The kinematical cuts we adopt in this calculation are~\cite{Alekseev:2010rw}
\begin{align}
&0.004<x<0.7,~~ 0.1<y<0.9,~~ z > 0.2,\nonumber\\
&P_T>0.1\,\textrm{GeV},~~Q^2>1\, \textrm{GeV}^2,\nonumber\\
&W>5\,\textrm{GeV}, ~~ E_h > 1.5\, \textrm{GeV}.
\end{align}

Similar to the case at HERMES and JLab, our theoretical prediction shows that the $g_T D_1$ term dominates the asymmetry $A_{\mathrm{LT}}^{\cos{\phi_S}}$, but the size of the asymmetry is less than $1\%$, which is clearly smaller than that at HERMES and JLab.
In the case of the asymmetry $A_{\mathrm{LT}}^{\cos{(2\phi_h -\phi_S)}}$, again we find that the main contribution is from the $g_T^\perp D_1$ term and the size is almost consistent with zero.
This is because the asymmetries we study appear at subleading twist, at which the effects will be suppressed by a factor of $1/Q$, and the $Q^2$ at COMPASS is larger than those at HERMES and JLab.
We note that our results for charged pions agree with the COMPASS preliminary measurements on the asymmetries $A_{\mathrm{LT}}^{\cos{\phi_S}}$ and $A_{\mathrm{LT}}^{\cos(2\phi_h -\phi_S)}$ for charged hadrons, within the current statistical accuracy~\cite{Parsamyan:2014uda}.

\section{Conclusion}
\label{conclusion}
In this work, we explored the roles of the twist-3 TMD distributions for the $u$ and $d$ valence quarks in the transverse target DSAs at subleading twist.
We calculated the $T$-even twist-3 TMD distributions $g_T$ and $g_T^\perp$, together with the $T$-odd and chiral-odd twist-3 TMD distributions $e_T$ and $e_T^\perp$, in a spectator model with both the scalar and axial-vector diquarks.
We distinguished the isoscalar ($ud$-like) and the isovector ($uu$-like) spectators for the axial-vector diquark and considered their differences in the calculation.
To generate $T$-odd structure, we employed the one-gluon exchange between the struck quark and the spectator; to obtain finite results, we chose the dipolar form factor for the nucleon-quark-diquark coupling. We also analyzed the flavor dependence of the four twist-3 TMD distributions as functions of $x$ and $k_T$, respectively.

By employing the model results on the distributions under the TMD framework, we predicted the transverse target DSAs $A_{\mathrm{LT}}^{\cos\phi_S}$ and $A_{\mathrm{LT}}^{\cos(2\phi_h -\phi_S)}$ for $\pi^+$, $\pi^-$, and $\pi^0$ productions in SIDIS at the kinematics of HERMES, JLab, and COMPASS.
We find that the DSA $A_{\mathrm{LT}}^{\cos\phi_S}$ is large at the kinematics of JLab, and the DSA $A_{\mathrm{LT}}^{\cos(2\phi_h -\phi_S)}$ is sizable at JLab. These two DSAs are not negligible at HERMES.
Furthermore, the comparison between different origins of the asymmetries shows that the $T$-even twist-3 TMD distributions $g_T$ and $g_T^\perp$ play an important role in these asymmetries.
Particularly, for the $\cos\phi_S$ asymmetry, the dominative contribution is from the $g_T D_1$ term; for the $\cos(2\phi_h -\phi_S)$ asymmetry, the main contribution is from the  $g_T^\perp D_1$ term.
The $e_T H_1^\perp$ and $e_T^\perp H_1^\perp$ terms almost give negligible contribution, except for
the asymmetry for $\pi^+$ and $\pi^-$ production at higher $P_T$.

Based on the above discussion, we conclude that sizable transverse double spin asymmetries may be accessible at the kinematics of HERMES and JLab, by performing the SIDIS experiments with transverse polarized nucleon target or analyzing the available data, although the effects might not be observable at COMPASS.
Moreover, the measurements on the $\cos\phi_S$ and $\cos(2\phi_h -\phi_S)$ asymmetries may be employed to obtain information of the $T$-even twist-3 distributions $g_T$ and $g_T^\perp$, since the contributions from $e_T$ and $e_T^\perp$ are negligible.
In particular, SIDIS provides a unique opportunity to probe $g_T^\perp$, since $g_T^\perp$ decouples from inclusive DIS.
Future comparisons with experimental data on these effects can provide more clear probes on the structure of the nuclear and deepen our understanding on the roles of the twist-3 TMD distributions in transverse spin asymmetries.

\section*{Acknowledgements}
W. M. thanks Tianbo Liu for the valuable discussion and inspiration.
This work is partially supported by the National Natural Science
Foundation of China (Grants No.~11120101004, No.~11005018, and No.~11035003), by the Qing Lan Project (China), and by Fondecyt (Chile) Grant No.~1140390.
W.M. is supported by the Scientific Research Foundation of the Graduate School of SEU (Grant No.~YBJJ1336).
Z. L. is grateful to the hospitality of Universidad T\'ecnica Federico Santa Mar\'\i a where part of this work was finished.

\end{document}